\documentclass[12pt,a4paper]{article}

\pdfoutput=1
\usepackage{amsmath}
\usepackage{amsfonts}
\usepackage{amssymb}
\usepackage{graphicx}
\usepackage{array}
\usepackage{cancel}
\usepackage{caption}
\usepackage{wrapfig}
\usepackage{mathrsfs}
\usepackage{cancel}
\usepackage{siunitx}
\usepackage{amsthm}
\usepackage{tikz-cd}
\usetikzlibrary{babel}

\usepackage[left=1.5cm,right=1.5cm,top=1.5cm,bottom=1.5cm,includefoot,footskip=1.5cm]{geometry}
\usepackage[OT1, T2A]{fontenc}
\usepackage[utf8]{inputenc}
\usepackage[english]{babel}
\usepackage[toc]{appendix}
\usepackage{csquotes}

\usepackage[sorting=none, citestyle=numeric-comp,
backend=biber
]{biblatex}
\usepackage{biblatex2bibitem}

\usepackage[colorlinks=true,
linkcolor=blue,citecolor=green,breaklinks=true]{hyperref}

\let\{=\lbrace
\let\}=\rbrace
\let\[=\lbracket
\let\]=\rbracket

\let\ep = \epsilon
\let\w = \omega

\newcommand\ili{\int\limits}
\newcommand {\lb} {\left(}
\newcommand {\rb} {\right)}

\newcommand\p{\partial}

\newcommand\const{\text{const}}

\newcommand{\bl}{\left\langle}
\newcommand{\br}{\right\rangle}

\newcommand{\mt}[1]{\mathcal{#1}}
\newcommand{\mtf}[1]{\mathfrak{#1}}

\DeclareMathOperator{\sign}{sign}

\numberwithin{equation}{section}
\usepackage{authblk}
\author[1,2]{E.T.Akhmedov}
\author[3]{I.V.Kochergin}
\author[1]{M.N.Milovanova}
\affil[1]{Moscow Institute of Physics and Technology, Institutskii per. 9, 141700, Dolgoprudny, Russia}
\affil[2]{Institute for Theoretical and Experimental Physics, B. Cheremushkinskaya 25, 117218, Moscow, Russia}
\affil[3]{Kharkevich Institute for Information Transmission Problems, 127994, Moscow, Russia}

\title{Isometry invariance of exact correlation functions in various charts of Minkowski and de Sitter spaces}

\begin{document}

\maketitle

\begin{abstract}
We consider quantum field theory with selfinteractions in various patches of Minkowski and de Sitter space-times. Namely, in Minkowski space-time we consider separately right (left) Rindler wedge, past wedge and future wedge. In de Sitter space-time we consider expanding Poincare patch, static patch, contracting Poincare patch and global de Sitter itself. In all cases we restrict our considerations to the isometry invariant states leading to maximally analytic propagators. We prove that loop corrections in right (left) Rindler wedge, in the past wedge (of Minkowski space-time), in the static patch and in the expanding Poincare patch (of de Sitter space-time) respect the isometries of the corresponding symmetric space-times. All these facts are related to the causality and analyticity properties of the propagators for the states that we consider. At the same time in the future wedge, in the contracting Poincare patch and in global de Sitter space-time infrared effects violate the isometries.  
\end{abstract}
\newpage

\section{Introduction}

In the study of various phenomena in strong background gravitational fields one frequently has to consider quantum field theory within patches (aka regions or wedges, or charts) of entire space-times rather than within whole space-times themselves. E.g. in the study of the Unruh effect \cite{Unruh:1976db} one considers the right Rindler wedge, which is a quarter of the entire Minkowski space-time. At the same time, in the study of inflation \cite{Starobinsky:1980te,PhysRevD.23.347,PhysRevLett.48.1220,Linde:1981mu} one restricts the consideration to the expanding Poincare patch, which is a half of the entire de Sitter space-time \cite{Starobinsky:1980te,PhysRevD.23.347,PhysRevLett.48.1220,Linde:1981mu}.

Unruh effect originates from the fact that Poincare invariant state (so called Minkowski vacuum) is seen as the thermal state (i.e. with the planckian distribution for the exact modes) from the point of view of the Rindler (eternally accelerated) observer. In the study of the inflation one also frequently restricts the consideration to the de Sitter isometry invariant states or to such states which inevitably rapidly approach the invariant states \cite{Bunch:1978yq,Mottola:1984ar,Allen:1985ux}. See, however, e.g. \cite{Akhmedov:2021rhq} for an alternative proposition: to check the destiny of a generic Hadamard states under the time evolution and see whether equilibration happens first or the backreaction leads to a strong deformation of the geometry and only then the equilibration happens.

For the isometry invariant states the tree-level propagators are functions of the invariants (geodesic distances and signs of the time differences) rather than functions of each their arguments separately. However, in the loops, if one restricts the integration over vertices to a patch of entire space-time, the correction naively seems to violate the isometry invariance. In fact, the measure of integration associated with vertices violates the invariance, because there are generators of the symmetry group of the entire space-time which move the patch.

Our goal here is to address the question of the isometry invariance of the loop corrections to correlation functions in various patches of Minkowski and de Sitter space-times. Such a question was already addressed by several authors in the right Rindler wedge and in the expanding Poincare patch \cite{Hollands:2010pr,Higuchi:2010xt,Higuchi:2020swc}. We extend those considerations to the future (or upper) and past (or lower) wedges of Minkowski space-time, to the contracting Poincare patch of the global de Sitter space-time and to the global de Sitter itself. We also use a different way of reasoning and by product prove several statements concerning the Schwinger-Keldysh and Feynman techniques in the patches and in general situations.

The deeper reason why we consider quantum field theory in such an unusual circumstances is as follows. We still do not really know what was the initial state of our Universe at the start of the inflation (if the inflationary epoch did really take place). Namely we do not know the geometry of the initial Cauchy surface, the basis of modes and even whether one should use for the quantization the basis in the entire space-time or on the initial Cauchy surface only. It is very hard to address such a problem in full generality. To keep task small it is meaningful to consider various initial Hadamard states (including initial Cauchy surfaces of different geometry) and start from simple symmetric situations. In this note we show that even for such situation the quantum field theory dynamics strongly depends on the geometry of the initial Cauchy surface and infrared effects play the key role. At the same time we study the dynamics in various patches of Minkowski space-time just as a simple training example. 

In the standard cases (right Rindler wedge and expanding Poincare patch) we confirm the previously made observations. Namely, we show that for the maximally analytic invariant propagators loop corrections respect the isometry. However, in some of the new cases (in the future wedge, in the contracting Poincare patch and in the global de Sitter space-time) we find that loop corrections violate the corresponding isometry symmetry due to infrared effects. We summarize our results and clarify their physical meaning in the concluding section.

\section{Right Rindler wedge}\label{RinWed}
In this section we consider interacting massive real scalar field theory in the $d$-dimensional Rindler wedge of Minkowski space-time:

\begin{equation}
    S = \int d^dX \, \sqrt{|g|} \, \left[\frac12 \left(\partial_\mu \phi\right)^2 - \frac{m^2}{2} \phi^2 - \frac{\lambda}{n!} \, \phi^n\right].
\end{equation}
Here $n \geq 3$ and in our study the possible non-renormalizability or stability of this theory is not important. We are interested in the properties of the loop corrected correlation functions under the action of the Poincare symmetry. While a similar question was addressed in \cite{Higuchi:2020swc}, we will present a different analysis which does not use the momentum space and, hence, can be generalised to curved spaces. Besides that, using the same method we will also consider other charts of the Minkowski space-time and of the de Sitter space-time as well. 

The relation between Minkowskian, $X^\mu$, $\mu = 0,\dots, d-1$, and the Rindler right wedge, $X^1 \geq \left|X^0\right|$, coordinates is as follows:
\begin{equation}
    X^0 = e^\xi \sinh\tau,\quad X^1 = e^\xi \cosh\tau,
\end{equation}
the other $d-2$ coordinates $X^a$, $a=2,\dots,d-1$ are unchanged. We use the standard euclidean metric for them. The corresponding Rindler metric induced from the Minkowskian one is
\begin{equation}
ds^2 = e^{2\xi}(d\tau^2-d\xi^2) - (dX^a)^2.\label{rinmetr}
\end{equation}
We consider the Rindler thermal state with the inverse temperature $\beta = 2\pi$ in units of acceleration \cite{Unruh:1976db, Crispino:2007eb,Troost:1978yk,Hollands:2014eia,Moschella:2009ub,Akhmedov:2020ryq}. Namely in the Feymnan and Wightman propagators,
\begin{equation}
    F(X,X') = \bl T \phi(X) \phi(X') \br_0,\quad W(X,X') = \bl \phi(X) \phi(X')\br_0,\label{Fprop}
\end{equation}
correspondingly, the expectation value is taken over the state that respects the Poincare isometry of the flat space-time. As a result, for such a state $F$ and $W$ are functions of the geodesic distance between $X$ and $X'$ and of the sign of the time difference rather than of each their argument separately. This is the standard state to consider in high energy particle physics.

The correlator with the reverse time-ordering, as well as the Wightman function with swapped points can be expressed via (\ref{Fprop}) by the complex conjugation:
\begin{equation}
    \bar{F}(X,X') = \bl \bar{T} \phi(X) \phi(X') \br_0,\quad W(X',X) = \bar{W}(X,X').\label{Fprop1}
\end{equation}
For the Poincare invariant state these propagators are also functions of the geodesic distance and of the sign of the time difference.

\subsection{The Feynman perturbation theory does not work}

Let us consider for the beginning the standard Feynman's diagrammatic technique in the right Rindler wedge, $X^1 \geq \left|X^0\right|$, with the propagator given by (\ref{Fprop}). Our goal is to show that it gives an incorrect result in the circumstances under consideration. In the next subsection we will show that the Schwinger-Keldysh technique provides the correct result.

Consider a Feynman diagram and an internal vertex $Y$ in it connected with $n$ external points $X_1,\dots,X_n$. In x--space the contribution of this vertex to the diagram is given by the following integral:
\begin{equation}
    I(X_1,\dots,X_n) = \ili d^d Y \theta(Y^1-Y^0)\theta(Y^1+Y^0)\prod_{j=1}^n F(Y,X_j).
\end{equation}
The $\theta$-functions restrict the integration region to the right wedge, we also assume that $X_i$ belong to the wedge as well. 

We wish to check whether $I$ is Poincare-invariant or not, i.e. does the result of the integration over the internal vertex $Y$ in the last equation depend only on invariant quantities such as geodesic distances and signs of the time differences or not. 

To do the check one can calculate the variation of $I$ under an infinitesimal coordinate transformation. For instance, consider the following translation, which moves the right wedge:
\begin{equation}
Y^1 \to Y^1 + a.\label{trans}
\end{equation}
The coordinates $X_i$, $i = 1,\dots, n$ are changed as well accordingly, such that the Feynman propagators under the integral do not change. The whole integral then changes as follows:
\begin{equation}
    \delta_a I(X_1,\dots,X_n) = a\ili d^d Y \Big[\delta(Y^1- Y^0)\theta(Y^1+ Y^0) + \delta(Y^1+Y^0)\theta(Y^1-Y^0) \Big] \prod_{j=1}^n F(Y,X_j).\label{delI1}
\end{equation}
To proceed, we need to take into account that $F(Y,X_j)$ are functions of the intervals $(Y-X_j)^2$ and are defined as the lower boundary value of an analytic function in the lower half-plane:
\begin{equation}
    F(Y,X_j) = \mathfrak{F}[(Y-X_j)^2 - i\ep].\label{Fiep}
\end{equation}
Explicit form of this function is not relevant for our considerations, but it is not hard to find out that in flat space-time it is proportional to one of the Hankell functions.

{It is also important to note that strictly speaking $F(Y,X_j)$ is a distribution rather than a function, and the limit $\ep \to 0$ should be taken in a distributional sense (see \cite{Kravchuk:2020scc} for a review). Besides that, this limit is defined in the complex plane of time coordinate, which we will discuss in more detail in section \ref{EuCont} (see e.g. fig. \ref{PropCuts}). Hence, in all subsequent expressions terms with $i\ep$-prescriptions should be understood as such distributional limits. Note that outside of singular points at the light cone they nevertheless admit representations in terms of ordinary functions. Furthermore, the Pauli-Villars regularization, which we discuss in appendix \ref{PVreg}, removes all the singularities and the propagator becomes well-defined everywhere on the real line.}

The interval itself can be expressed as follows:
\begin{equation}
(Y-X_j)^2 = (Y^0-Y^1)(Y^0+Y^1) - (Y^0-Y^1)(X_j^0+X_j^1) - (Y^0+Y^1)(X_j^0-X_j^1) - (Y^a)^2 + 2 X_j^a Y^a + (X_j)^2.
\end{equation}
Taking the integral over $Y^1$ in (\ref{delI1}) we find:
\begin{equation}
\begin{aligned}
    \delta_a I(X_1,\dots,X_n) &= a\ili d^{d-1} Y \theta(Y^0)\left[ \prod_{j=1}^n  \mathfrak{F}\Big[  (X_j^1-X_j^0)(2Y^0 - X_j^1 - X_j^0 -i\ep) - (Y^a - X_j^a)^2 \Big] \right. +\\
    &+ \left.  \prod_{j=1}^n \mathfrak{F}\Big[ (X^1_j + X^0_j) (2Y^0 -X_j^1 + X_j^0-i\ep) - (Y^a-X^a_j)^2 \Big] \right],\label{delIF}
\end{aligned}
\end{equation}
where $d^{d-1} Y = dY^0 dY^2\dots dY^{d-1}$ and we used that $X_j^1 + X_j^0 > 0$, $X_j^1 - X_j^0 > 0$, because all points that we consider here are in the right wedge. Also, we have changed $-Y^0 \to Y^0$ in the second term under the integral. The integrand is a lower boundary value of an analytic function of $Y^0$, but due to the presence of $\theta\left(Y^0\right)$ the contour of integration cannot be closed (unlike e.g. the case considered in \cite{Akhmedov:2020jsi} for the Feynman technique in the Poincare region of the anti de Sitter space-time). Also note that the two terms in the brackets cannot cancel each other, as the first one is a function of $X_j^1 - X_j^0$, while the second one is a function of $X_j^1 + X_j^0$. 

In all, $I(X_1,\dots,X_n)$ is not Poincare-invariant, because it changes under the action of the generators of the corresponding group. This is just a revelation of the fact, as we discuss in the next subsection, that the standard stationary perturbation theory (Feynman digrammatic technique) is not applicable in the case under consideration. One has to apply the Schwinger-Keldysh perturbation theory.

\subsection{The Schwinger-Keldysh time contour}\label{skc}

As was mentioned above, we consider a thermal state in Rindler wedge of the Minkowski space-time. Hence, it is appropriate to use the Schwinger-Keldysh diagrammatic technique rather than the Feynman one. That is the reason why we have found problems with loop corrections in the previous subsection. Similar observations for de Sitter space were made in e.g. \cite{Higuchi:2008tn,Higuchi:2009ew}.

In the Schwinger-Keldysh technique one has to use the double time contour going from the past infinity (in the case of an equilibrium state) to the future infinity and then backwards \cite{Kamenev, Kamenev2, LL10}. As the result, one should assign to each internal vertex either <<$+$>> or <<$-$>>, depending to which part of the countour it belongs to, and use the following propagators to compute diagrams:
\begin{equation}
G_{--}(X,Y) = F(X,Y),\quad G_{+-} = W(X,Y),\quad G_{-+} = W(Y,X),\quad G_{++}(X,Y) = \bar{F}(X,Y),\label{4props}
\end{equation}
where $F$ and $W$ are the two--point functions which were defined at the beginning of this section.

To calculate the contribution of a diagram one has to take the sum over the signs of internal vertices. Also, each <<$-$>> vertex has an additional $-1$ multiplier. Let us consider a part of a diagram in which the external vertices $X_1,\dots,X_k$ are of <<$-$>> type, while the rest of them ($X_{k+1},\dots,X_n$) are of <<$+$>> type. Let all these vertices be connected to an internal vertex $Y$. Then the contribution of the part of the diagram under consideration corresponds to the integral over the vertex $Y$ with the summation over its signs and has the following form (up to a common factor $(-1)^k$):
\begin{equation}
\begin{aligned}
I_K(X_1,\dots,X_n) = \ili d^d Y \theta(Y^1-Y^0)\theta(Y^1+Y^0) &\left[\prod_{j=1}^k F(Y,X_j) \prod_{j=k+1}^n \bar{W}(Y,X_j) - \right.\\
&-\left.\prod_{j=1}^k W(Y,X_j) \prod_{j=k+1}^n\bar{F}(Y,X_j) \right].
\end{aligned}\label{IK}
\end{equation}
The loop integrals of course contain standard UV divergences which have to be regularized. We assume the Pauli-Villars regularization scheme, in which from the propagators defined in (\ref{Fprop}) and (\ref{Fprop1}) one subtracts a sufficient number of propagators of massive fields to get a finite expression for the diagram. For our further considerations it is important that the expressions we use in the diagrams instead of $F$ and $W$ have exactly the same analytic properties as functions of geodesic distances. Thus, below we assume the Pauli-Villars regularization scheme, but use the same notations $F$ and $W$ for the subtracted two-point functions. More details are provided in the appendix \ref{PVreg} for the case of the de Sitter space-time.

Also note that the expression for $I_K$ does not include tadpoles. However, they will not change any of our considerations. Namely, the contribution of a tadpole connecting $Y$ with itself to the integrand of (\ref{IK}) consists of an additional factor $F(Y,Y)$ in the first summand and $\bar{F}(Y,Y)$ in the second one. These functions are constants due to Lorentz-symmetry, which become finite after the regularization. To ensure the cancellation of vacuum bubbles the constants $F(Y,Y)$ and $\bar{F}(Y,Y)$ should coincide\footnote{Note that this condition can be violated only in a non-stationary situation for the in-out Feynman propagator. But we consider here in-in Feynman propagators only (see e.g. \cite{Akhmedov:2009ta,Akhmedov:2019esv} for a related discussion in de Sitter space-time).}, i.e. $\text{Im}\,F(Y,Y) = 0$. It means that the contribution of tadpoles to the whole diagram is just a constant factor, which is irrelevant for our considerations.

Let us consider now the transformation of $I_K(X_1,\dots,X_n)$ under the action of the same translation (\ref{trans}) as in the previous subsection. We will show now that $\delta_\epsilon I_K(X_1,\dots,X_n) = 0$. The same can be shown for other generators of the Poincare group. I.e. by showing that $I_K(X_1,\dots,X_n)$ is invariant we prove that it is a function of such isometry invariants as geodesic distances and signs of time differences. 

To start the proof we should express the Wightman function in terms of the boundary value of a complex function $\mtf{F}$ of a complex variable as follows:
\begin{equation}
W(Y,X) = \mathfrak{F}\Big[ (Y-X)^2 - i\ep \sign(Y^0-X^0) \Big].\label{Wiep}
\end{equation}
Here we use the same function as the one that we used to express the Feynman in (\ref{Fiep}).
Explicit form of this function is not relevant for our considerations. It is important that it is an analytic function on the cutted complex plane of the geodesic distance. The cut is going along such values of the geodesic distance, which corresponds to time-like separations between $X$ and $Y$ and starts with the light-cone separation. As a result $\text{Im}\,\mtf{F}\neq 0$ on the cut and the commutator of the field operators $\phi(X)$ and $\phi(Y)$ is not zero for time-like separations, as it should be on the physical grounds. In fact, this way one obtains a non-zero classical retarded propagator in the field theory under consideration. At the same time the function $\mtf{F}$ is real for space-like separations. That is necessary to fulfil the condition of causality and to have the commutator of the field operators zero for spatial separations. Finally the written above $i\ep$-prescriptions in the expressions for $W$ and $F$ just specifies which side of the cut should be taken to define the propagator for time-like separations.

To proceed with the proof it is convenient to introduce the following notations for the squares of geodesic distances if $Y$ is on the boundary of the wedge :
\begin{equation}
\begin{aligned}
d_1(Y,X_j) &= (X_j^1- X_j^0)(2Y^0-X_j^1-X_j^0) - (Y^a-X_j^a)^2,\\
d_2(Y,X_j) &= (X_j^1+ X_j^0)(2Y^0-X_j^1+X_j^0) - (Y^a-X_j^a)^2.
\end{aligned}\label{d12}
\end{equation}
In the first expression we assume that $Y^0 = Y^1$ and in the second one that $Y^0 = -Y^1$. We also changed $Y^0 \to - Y^0$ in $d_2$ for future convenience. Then we find that the transformation of $I_k$ under the translation (\ref{trans}) can be written as:
\begin{equation}
\delta_a I_K(X_1,\dots,X_n) = a(\delta^{(1)} I_K + \delta^{(2)} I_K),\label{delIK}
\end{equation}
where
\begin{equation}
\begin{aligned}
\delta^{(1)} I_K &= \ili d^{d-1} Y \theta(Y^0)\left\{ \prod_{j=1}^k\mathfrak{F}\Big[d_1(X_j,Y) - i\ep\Big] \prod_{j=k+1}^n \mathfrak{F}\Big[d_1(X_j,Y) + i\ep \sign(Y^0 - X_j^0)\Big]-\right.\\
&-\left. \prod_{j=1}^k\mathfrak{F}\Big[d_1(X_j,Y) - i\ep \sign(Y^0-X^0_j)\Big] \prod_{j=k+1}^n \mathfrak{F}\Big[d_1(X_j,Y) + i\ep\Big]\right\}, \quad {\rm and}\\
\delta^{(2)} I_K &= \ili d^{d-1} Y \theta(Y^0)\left\{ \prod_{j=1}^k\mathfrak{F}\Big[d_2(X_j,Y) - i\ep\Big] \prod_{j=k+1}^n \mathfrak{F}\Big[d_2(X_j,Y) - i\ep \sign(Y^0 +X_j^0)\Big]-\right.\\
&-\left. \prod_{j=1}^k\mathfrak{F}\Big[d_2(X_j,Y) + i\ep \sign(Y^0+X^0_j)\Big] \prod_{j=k+1}^n \mathfrak{F}\Big[d_2(X_j,Y) + i\ep\Big]\right\}.\\
\end{aligned}\label{Del1Del2}
\end{equation}
Let us start with the consideration of $\delta^{(1)} I_K$. Note that the integrand in the corresponding expression vanishes, when $Y_0$ is larger than all $X^0_j$. In fact, in such a case the $i\ep$-prescriptions in the first and second terms under the integral coincide, because $\sign(Y^0 -X_j^0) = 1$. Hence, when $Y_0$ is larger than all $X^0_j$ we find that $\delta^{(1)} I_K = 0$. 

Now assume that $Y^0<X_j^0$ for some $j$, then we have\footnote{We can use strict inequalities as the sets of zero measure obviously do not contribute to the integrals.}:
\begin{equation}
X_j^0 > Y^0 > 0,
\end{equation}
where the restriction $Y^0 > 0$ appears due to the theta-function under the integral in (\ref{Del1Del2}). From (\ref{d12}) we see that then $d_1(Y,X_j) < 0$, because inside the Rindler wedge $X_j^1 > X_j^0$. Therefore the interval between $X_j$ and $Y$ is space-like, so $i\ep$-prescription does not matter, because the function $\mtf{F}$ is real for such an argument: for space-like separations all four propagators from (\ref{4props}) with $X=X_j$ coincide. Hence, the integrand in $\delta^{(1)} I_K$ vanishes and again we find that $\delta^{(1)}I_K = 0$. This essentially completes the proof that for the transformation in question $\delta^{(1)} I_K$ is always zero due to the analytic properties of the propagators. 

\begin{figure}[ht!]
\centering\includegraphics[width=0.75\textwidth]{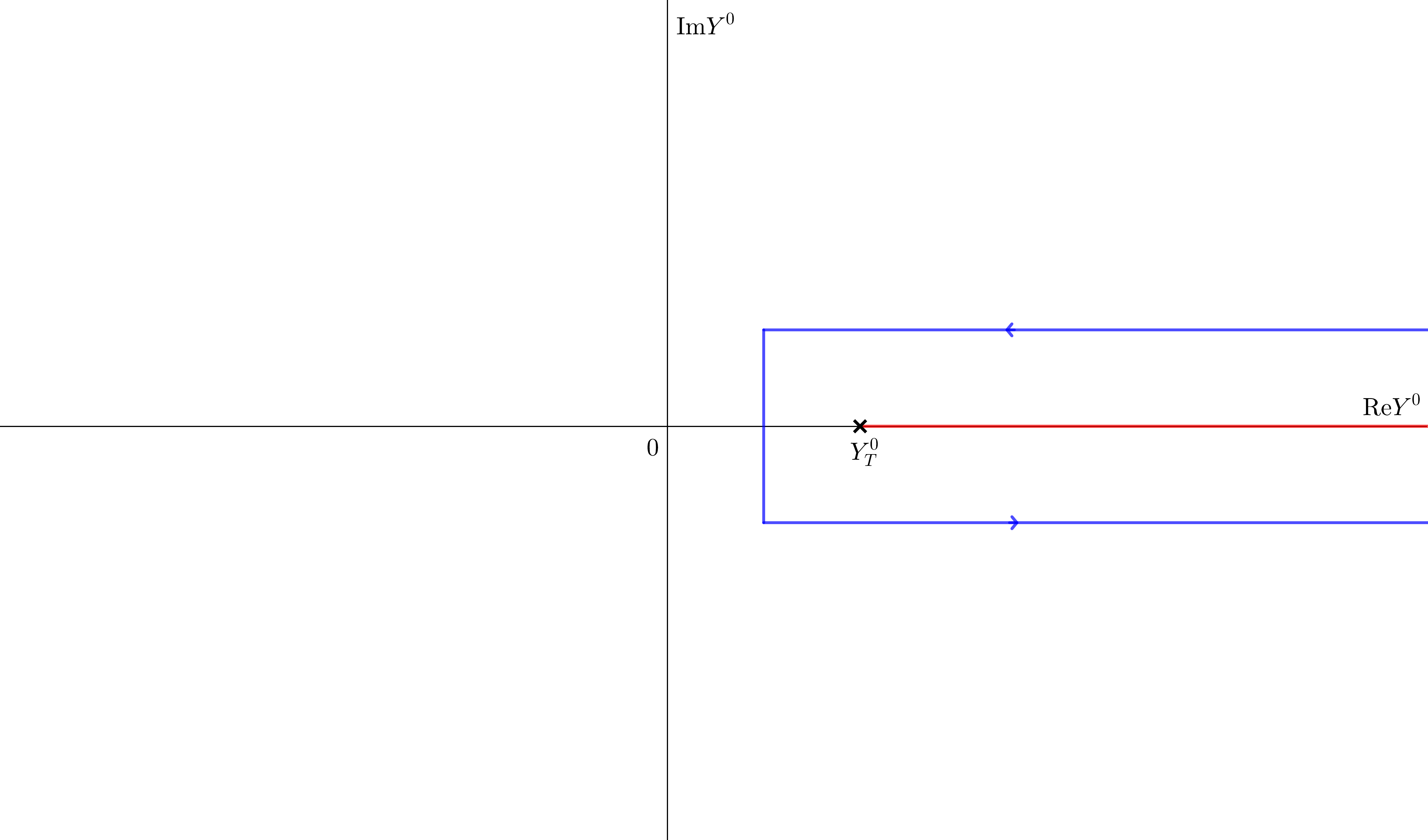}
\caption{The integration contour for $\delta^{(2)}I_K$ (blue), the branch cut is shown in red.} \label{I2Cont}
\end{figure} 

The case of $\delta^{(2)} I_K$ is slightly more complicated. The signs of $i\ep$-prescriptions in two terms in $V$ coincide if $\sign(Y^0+X_j^0) = -1$, i.e. when
\begin{equation}
X_j^0 < -Y^0 < 0.\label{ineqEp}
\end{equation}
(The second inequality again appears due to the presence of the step function under the corresponding integral for $\delta^{(2)} I_K$.) In such a case the integrand of $\delta^{(2)} I_k$ obviously vanishes. 

From (\ref{ineqEp}) it also follows that $d_2(Y,X_j) < 0$, i.e. the interval is space-like if $\sign(X^0+X_j^0) = -1$. It means that if the interval is time-like, $\sign(Y^0 + X_j^0) = 1$ and the $i \ep$-prescriptions in two terms in $\delta^{(2)} I_K$ are different~--- it is the opposite situation to what we have observed for the case of $\delta^{(1)} I_K$. As $i\ep$-prescription is irrelevant when the interval is space-like, we can replace $\sign(Y^0 + X_j^0)$ with $1$ for all $X_j^0$ and obtain that:
\begin{equation}
\delta^{(2)} I_k = \ili d^{d-1} Y \theta(Y^0) \left\{ \prod_{j=1}^{n} \mathfrak{F}\Big[d_2(X_j,Y) - i\ep\Big] -  \prod_{j=1}^{n} \mathfrak{F}\Big[d_2(X_j,Y) + i\ep\Big] \right\}.\label{delIK2}
\end{equation}
Now let us consider the properties of the product $\prod_{j=1}^n\mt{F}\Big[d_2(X_j,Y)\Big]$ as the function on the complex plane of $Y^0$. It is straightforward to see from (\ref{d12}) that $d_2(X_j,Y) < 0$ if $Y^0 \le 0$ (assuming that $X_j$ belongs to the right Rindler wedge). The interval first becomes light-like for some $Y^0_t > 0$, which is the lowest value of $Y^0$ such that $d_2(Y,X_j) = 0$ for some of the $j$'s. Hence, $\prod_{j=1}^n\mt{F}\Big[d_2(X_j,Y)\Big]$ has a cut along the halfline $(Y^0_t,+\infty)$, where $Y^0_t$ is the smallest solution of the equations $d_2(Y,X_j) = 0$. 

Now, similarly to (\ref{delIF}) we can attribute $i\ep$-prescription in (\ref{delIK2}) to the complex plane of $Y^0$ itself rather that to the complex plane of $d_2(X_j,Y)$. It means that the r.h.s. of (\ref{delIK2}) is nothing but the integral around the cut, as is shown on the Fig. \ref{I2Cont}. We can close this contour by a circle of the infinitely large radius going clockwise, on which the propagator vanishes.
As a result\footnote{It is worth stressing that for such manipulations with the contour we need only the convergence of $Y^0$ integrals, which is much weaker condition than the complete convergence.}, we obtain that $\delta^{(2)}I_k = 0$, because the integrand in (\ref{delIK2}) is analytical inside the closed contour.

In all, we have shown that $I_K(X_1,\dots,X_n)$ is invariant with respect to the translations along $X_1$. In the same way one can show that $I_K(X_1,\dots,X_n)$ remains intact under all other transformations of the Poincare group. 

Furthermore, note that by the translations along $X_1$ we can move the hyperplane $X^1 = X^0 = 0$, which is the edge of the right Rindler wedge, to $-\infty$. The loop integral does not change under such a move. It means that extending the integration region outside the Rindler wedge does not change the integral. Hence, the integration over the Rindler wedge inside the loop integrals in the Schwinger-Keldysh technique is equivalent to the integration over the entire Minkowski space-time. At the same time, the Schwinger-Keldysh technique in the whole Minkowski space-time for the Poincare invariant state provides the same answer for the loop integrals as the Feynman technique: the contribution of the backward going part of the time contour just cancels the vacuum bubble diagrams. We explain these points in greater details in the next subsections.

\subsection{Analytical continuation to the euclidean time contour}\label{EuCont}

There is another way to show that the Schwinger-Keldysh technique provides isometry-invariant expressions for the loop corrected correlation functions over the Poincare invariant state. Furthermore, this way directly shows that for the Poincare invariant state in the right Rindler wedge the Schwinger-Keldysh technique provides the same answer for the loop corrections as the Feynman technique in the entire Minkowski space-time.

This way is based on the analytical continuation to the euclidean signature directly from the Rindler wedge. Namely, note that the coordinate change $\tau = -i\kappa$ connects the right wedge (\ref{rinmetr}) with the euclidean space $\mathbb{R}^d$ with polar coordinates in the $(\kappa,\xi)$-plane, i.e. with the coordinate $\kappa$ being $2\pi$-periodic. In other words, this leads to the same result as the standard Wick rotation $X^0 = -iX^d$ of the whole Minkowski space-time. Furthermore, as we pointed out above, the Minkowski vacuum (Poincare invariant state) is a thermal state, when expressed via exact modes in Rindler coordinates. Hence, the analytically continued Feynman propagator $\mathfrak{F}$ (or Matsubara propagator) is $2\pi$-periodic in imaginary time. 

To perform the analytical continuation from the Rindler wedge to the Euclidean space in the loop integrals, we need to understand the analytic properties of the propagator $\mathfrak{F}$ in the complex plane of the time $\tau$. For simplicity let us consider the function $\mathfrak{F}[(X-Y)^2]$ when $X^0 = 0$, i.e. when $\tau_X = 0$, while $\tau = \tau_Y$. Also note that $\mathfrak{F}$ depends only on the difference $\tau_X-\tau_Y$ as the metric (\ref{rinmetr}) is $\tau$-independent, i.e. static.

\begin{figure}[ht!]
\centering\includegraphics[width=0.8\textwidth]{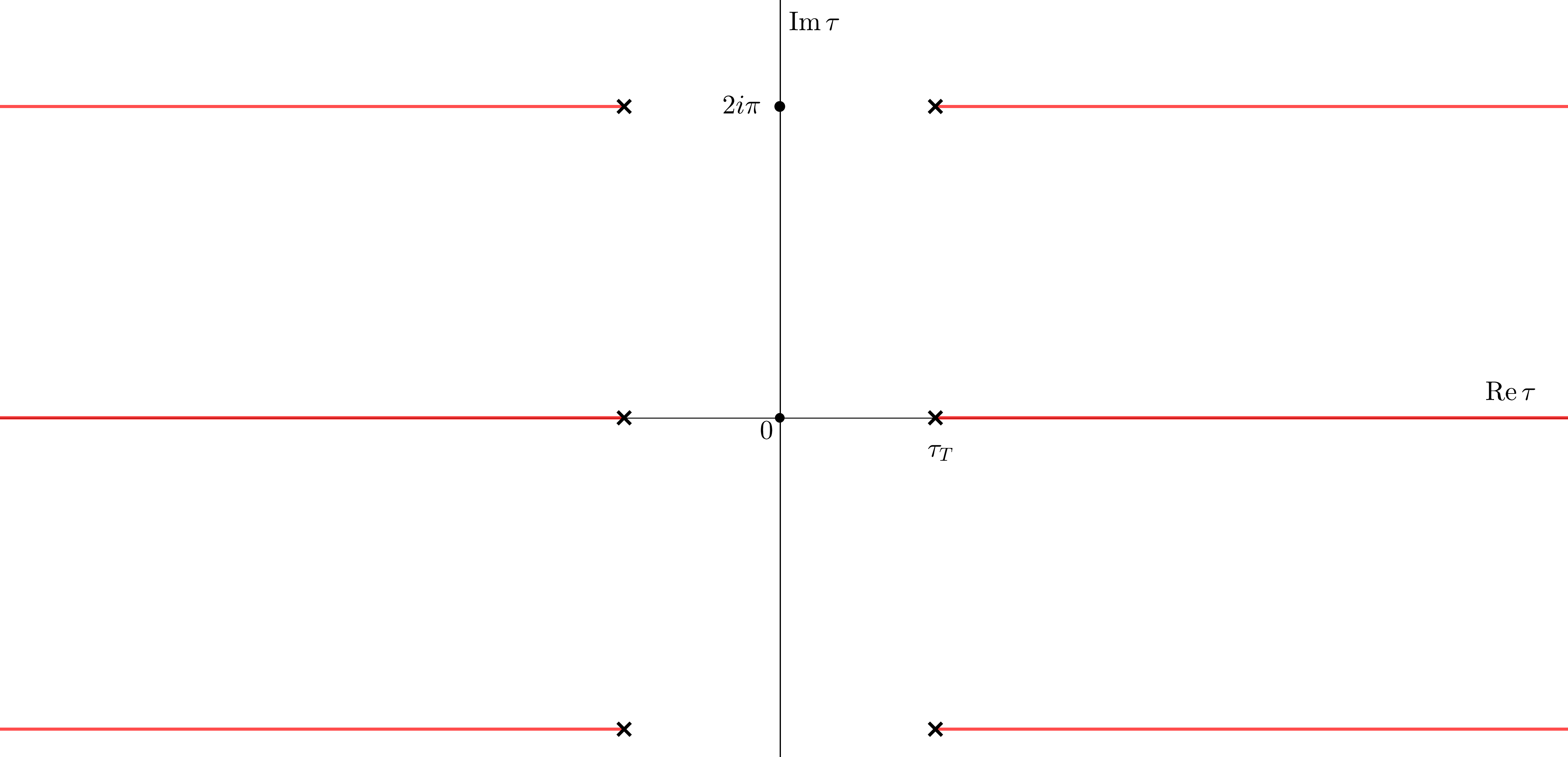}
\caption{The analytic structure of the thermal propagator in Rindler wedge. Red lines are the cuts in the complex $\tau$-plane.} \label{RinCuts}
\end{figure} 

The analytic structure of $\mathfrak{F}$ is shown on the Fig. \ref{RinCuts}. We assume that $\xi$ (the corresponding spatial Rindler coordinate of $Y$) and $Y^a$ are fixed, and $\tau_T$ is the time coordinate when the interval is light-like (with $\tau_T >0$). The propagator has the usual cuts along the lines $(\tau_T,\infty)$ and $(-\infty,-\tau_T)$, which correspond to the time-like separations between $Y$ and $0$. The $2\pi i$-periodicity means that $\mathfrak{F}$ also has additional branching points at $\pm \tau_T + 2\pi i k$, $k\in\mathbb{Z}$ with the corresponding cuts, as it is illustrated on the Fig. \ref{RinCuts}. Hence, the Feynman's integration contour over the $\tau$ (polar) coordinate in the loop integrals cannot be simply rotated to the imaginary axis as it is usually done in Cartesian coordinates\footnote{Such a simple rotation is performed when one does the analytical continuation in the Feynman's technique from the entire Minkowski space-time to the euclidean space in Cartesian coordinates.}. Thus, again we see that in the Rindler wedge loop corrections in the Feynman diagrammatic technique in x-space representation cannot be mapped to the integrals in the euclidean space. 

\begin{figure}[ht]
    \centering\includegraphics[width=0.6\textwidth]{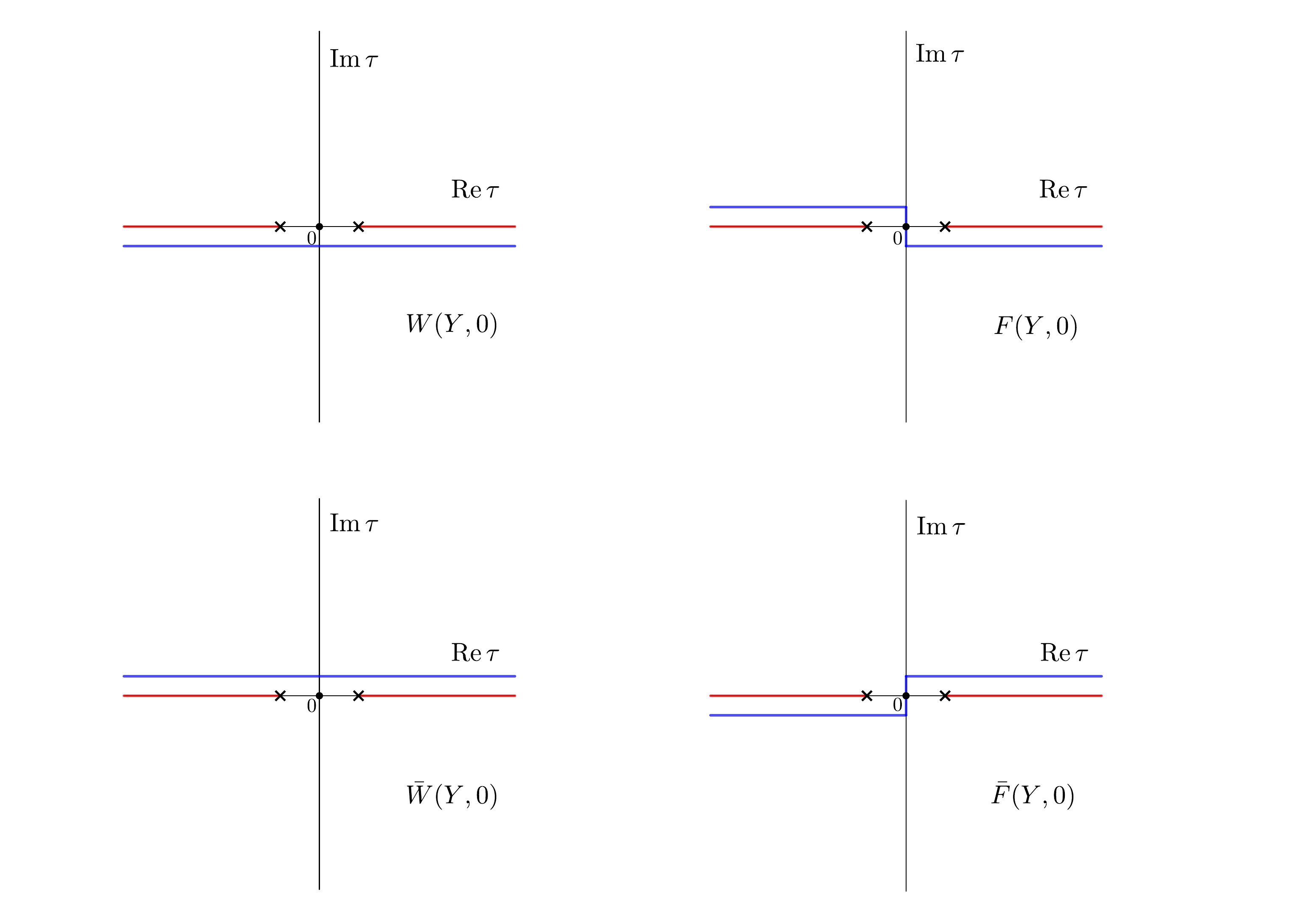}
    \caption{The propagators as boundary values; the red lines are the cuts; the blue lines depict the sides of the cut where the corresponding propagators are defined. We depict only those contours and cuts which are laying in the stripe Im $\tau \in (-\pi,\pi)$.}
    \label{PropCuts}
\end{figure}

Let us consider the situation in the case of the Schwinger-Keldysh time contour and of the corresponding diagrammatic technique. Our analysis will be more or less similar to the one presented in \cite{Higuchi:2010xt} for the static patch of the de Sitter space-time. Let us consider the same vertex integral $I_K$ as in (\ref{IK}) with $\tau$ being the Rindler time of the $Y$ coordinate. We need to represent it as a contour integral in the complex plane of $\tau$ rather than of $Y^0$. The propagators can be represented as boundary values of the analytic functions $\mtf{F}[(Y-X_j)^2]$ as in (\ref{Fiep}), (\ref{Wiep}). The cuts and contours are depicted on Fig. \ref{PropCuts} for the case when $\tau_j = 0$. Extra cuts coming from the periodicity in $\tau$ are omitted to simplify the picture. Note that the case with $\tau_j \neq 0$ can be simply obtained via the translation $\tau \to \tau + \tau_j$ ($\tau_j$ is a Rindler time of $X_j$) on the Fig. \ref{PropCuts}. In such a case the center of the line segment whose endpoints are branching points is shifted to $\tau_j$ from $0$.

The representation of eq. (\ref{IK}) in terms of a contour integral can be achieved if we represent its integrand as a boundary value of some analytic function similarly to the case of (\ref{delIK2}). It is convenient to split $I_K$ and its integrand into two parts:
\begin{equation}
\begin{aligned}
\hat{F}_n^- &= \prod_{j=1}^k F(Y,X_j) \prod_{j=k+1}^n \bar{W}(Y,X_j),\quad \hat{F}_n^+ = \prod_{j=1}^k W(Y,X_j) \prod_{j=k+1}^n\bar{F}(Y,X_j);\\
I_K^+&= \ili d^{d} y\,\sqrt{-g(y)} \hat{F}^+_n,\quad I_K^- = \ili d^{d} y\,\sqrt{-g(y)} \hat{F}^-_n,\label{HatFpm}
\end{aligned}
\end{equation}
where $y^\mu$ are the Rindler coordinates of $Y$ with $y^0 = \tau$ and $g_{\mu\nu}$ is the Rindler metric defined in (\ref{rinmetr}). Note that the determinant of the metric $g(y)$ is independent of $\tau$.

\begin{figure}[ht!]
\begin{minipage}[h]{1\linewidth}
\center{\includegraphics[width=0.8\textwidth]{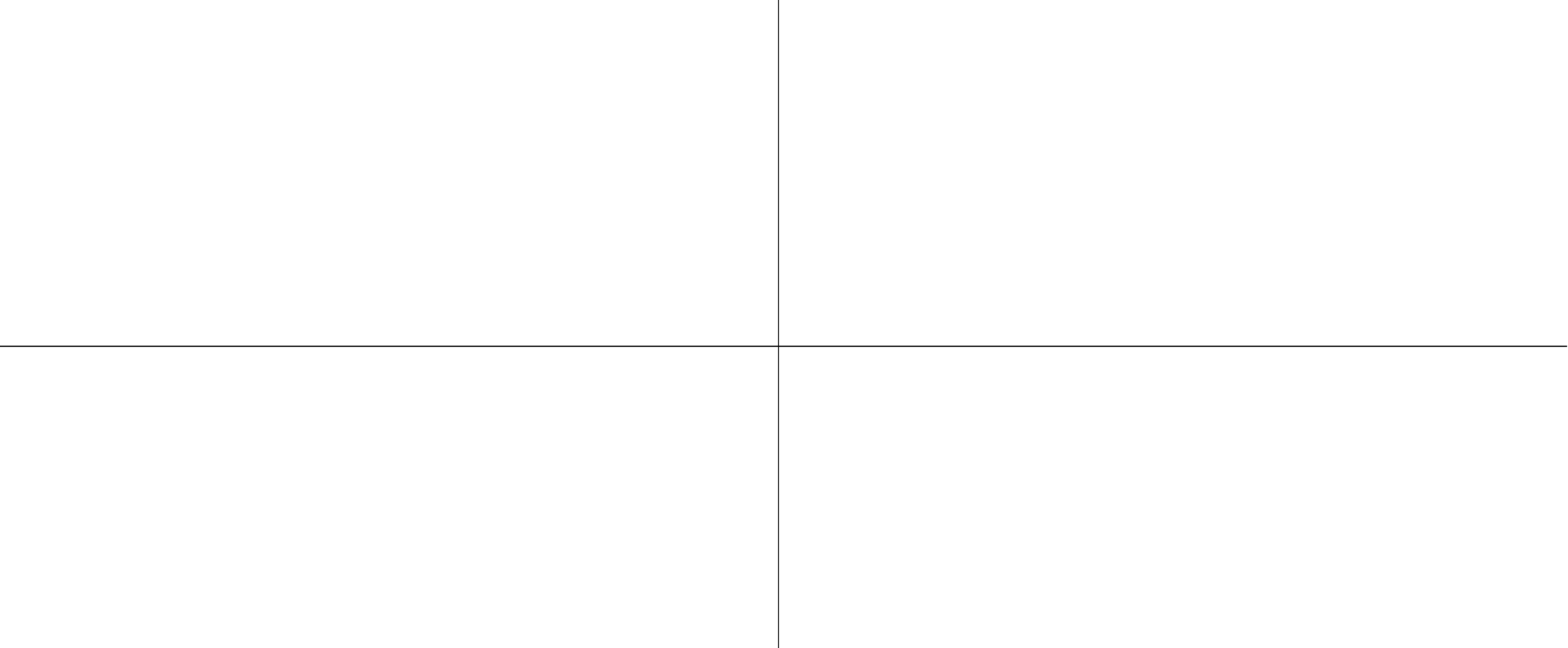} (a)}
\end{minipage}
\begin{minipage}[h]{1\linewidth}
\hfill

\hfill
\end{minipage}
\begin{minipage}[h]{1\linewidth}
\center{\includegraphics[width=0.8\textwidth]{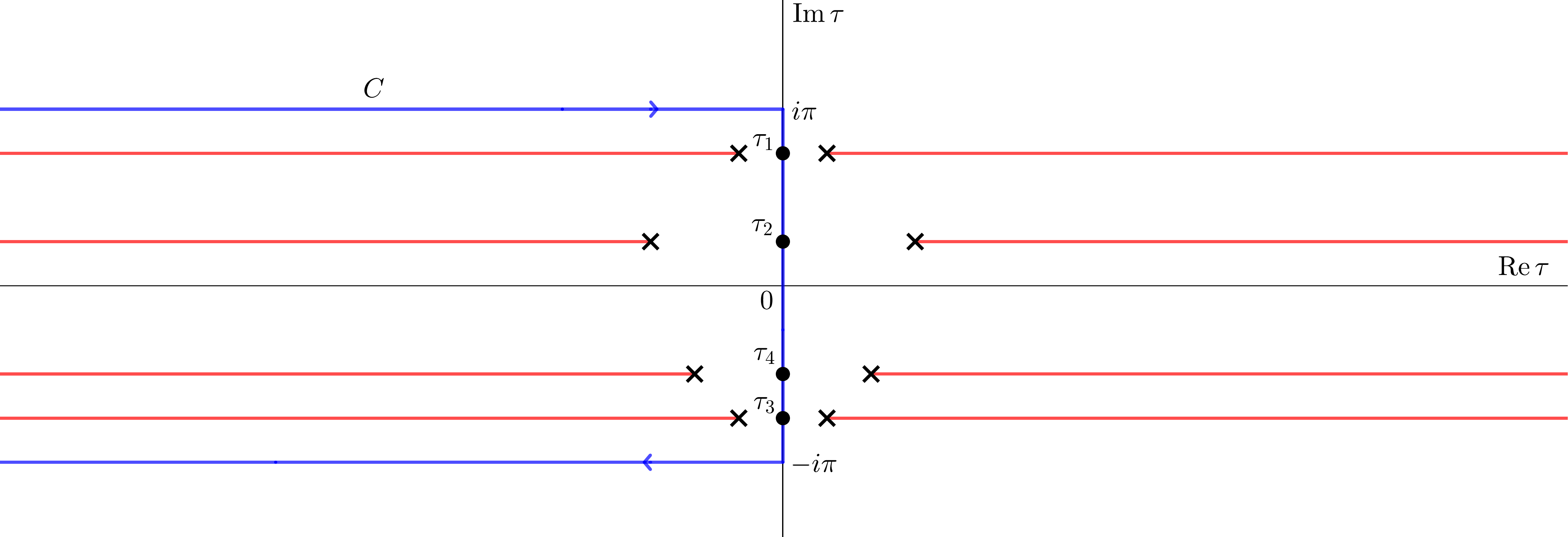} (b)}
\end{minipage}
\caption{An example of the contour in the loop integral when there are four external points. The contours of $\tau$ integration: (a) the points $\tau_j$ have small imaginary shifts, which we depict for clarity as finite shifts; the forward going part of the contour (along the real axis), $C_-$, corresponds to the first summand in (\ref{IK}), while the backwards going part, $C_+$, to the second one (its the negative sign is accounted for by the orientation w.r.t. $C_+$); (b) $\tau_j$ are analytically continued from the complex plane to the imaginary axis in the interval $(i\pi,\,-i\pi)$. The external points $X_1$ and $X_2$ have <<$-$>> sign while $X_3$ and $X_4$ have <<$+$>> sign.} \label{AnCont}
\end{figure} 

Next, one can define the following function $\mtf{F}_n(Y|X_1,\dots,X_n)$ on the complex $\tau$-plane :
\begin{equation}
\mtf{F}_n(Y|X_1,\dots,X_n) = \prod_{j=1}^n \mtf{F}[(Y-X_j)^2].
\end{equation}
Evidently the cuts of this function are defined as the union of the cuts of the functions in the product on the right hand side (RHS), which were described above. Of course the cuts associated with different $X_j$ overlap if all of $\tau_j$ are real. To avoid such an overlapping we can assign infinitesimal imaginary parts to $\tau_j$: $\text{Im}\tau_j = \ep_j$, where $\ep_j$ are pairwise distinct. Then $\hat{F}_n^+$ and $\hat{F}_n^-$ can be represented as their boundary values using the way of defining propagators via $\mtf{F}[(X_j-Y)^2]$ presented on the Fig. \ref{PropCuts}. Namely, according to (\ref{HatFpm}) we find that in the case of $\hat{F}^-$ we should take the lower value for the cuts associated with <<$-$>> vertices $X_j$ (e.g. $j \le k$) such that $\tau \ge \tau_j$ and the upper value for the rest of the cuts. Similarly, for $\hat{F}_n^+$ we should take the upper value for the cuts associated with <<$+$>> vertices $X_j$ (e.g. $j>k$) such that $\tau \ge \tau_j$ and the lower value for the rest of the cuts.

The integrals $I_K^-$ and $I_K^+$ from (\ref{HatFpm}) are contour integrals if the sides of the cuts where $\hat{F}^-_n$ and $\hat{F}^+_n$ are defined can be connected by a single curve which does not cross the cuts. The definition of such a contour leads to a condition on $\ep_j$: $\ep_i > \ep_j$ if one of the following conditions is satisfied:
\begin{itemize}
    \item $\tau_i < \tau_j$ and $X_i$, $X_j$ are <<$-$>> vertices;
    \item $\tau_i > \tau_j$ and $X_i$, $X_J$ are <<$+$>> vertices;
    \item $X_i$ is a <<$-$>> vertex and $X_j$ is a <<$+$>> vertex.
\end{itemize}
The possible choice is $\ep_j = \pm i\ep \, (1-\tanh \tau_j)$, where the <<$+$>> sign should be used for <<$-$>> vertices and vice versa. The contours $C^{\pm}$ for $I_K^{\pm}$ which go along the branch cuts as explained above can be defined via the same prescription $\tau \to \tau \pm i\ep \, (1-\tanh \tau)$, where the sign of the $i\ep$-term is opposite to the sign of the vertex $Y$. Note that $i\ep$-prescription for each vertex among $Y$ and $X_j$ depends only on the vertex itself, but not on their relative configuration. It means that we can apply the prescription to the whole diagram simultaneously rather than just to the subdiagram under consideration. This yields a representation of the whole diagram in terms of a multiple contour integral.

As $I_K = I_K^{-} - I_K^{+}$, it can be represented as an integral over the contour $C = C^{-} - C^{+}$. This contour is depicted on the Fig. \ref{AnCont} (a) for the case of two <<$-$>> points $X_1$, $X_2$ and two <<$+$>> points $X_3$, $X_4$. For convenience we deformed the contour in the region $\text{Re}\,\tau > \tau_j$ for all $j$, as the corresponding parts of $C^-$ and $C^+$ do not go around any cuts or poles. A similar contour for an expanding Poincare patch of de Sitter space is shown in \cite{Korai:2012fi}.

Now we can perform the analytical continuation to the euclidean theory using this contour integral representation. Due to $2\pi$-periodicity in imaginary time, we can assume that the time coordinates of the euclidean vertices belong to the line segment $(-i\pi, i\pi)$ along the imaginary time axis. To obtain the continuation of $I_K$ as a function of $X_j$, we need to move the time coordinates of $X_j$ to this line segment in such a way that the cuts do not cross each other. The contour $C$ then can be represented as a union of oriented line segments $(-\infty + i\pi,\, i\pi)$, $(i\pi,\,-i\pi)$ and $(-i\pi,\,-\infty-i\pi)$, where $\infty$ is the real infinity. The resulting position of the external points of the diagram and of the contour is depicted on the Fig. \ref{AnCont} (b). In particular, due to the $2\pi i$-periodicity the contributions from $(-\infty + i\pi,\, i\pi)$ and $(-i\pi,\,-\infty-i\pi)$ cancel each other, and we are left with the integration over the $(i\pi,-i\pi)$ segment along the imaginary time axis. This is precisely the correct euclidean contour with $\kappa \in (-\pi,\pi)$ for the euclidean or Matsubara diagrammatic technique. 

In all, the Schwinger-Keldysh technique in Rindler wedge can be analytically continued to the standard perturbation theory in euclidean space $\mathbb{R}^d$. This, once again, proves the Poincare invariance of loop corrected correlation functions for the thermal state with the canonical temperature ($\beta = 2\pi$ in the units of acceleration). All this is true despite the fact that one integrates in the loop integrals over the Rindler wedge in the vertices rather than over the entire Minkowski space-time. Besides that, our observations show that loop corrections in the Rindler wedge coincide with those in the Feynman technique calculated in the entire Minkowski space-time for the Poincare invariant state. In fact, the latter can be as well analytically continued to the euclidean space $\mathbb{R}^d$.

This observation completes our considerations of the right Rindler wedge.

\section{Causality of the Schwinger-Keldysh technique}\label{Caus}
\begin{figure}[ht!]
\begin{minipage}[h]{0.46\linewidth}
\center{\includegraphics[width=1\textwidth]{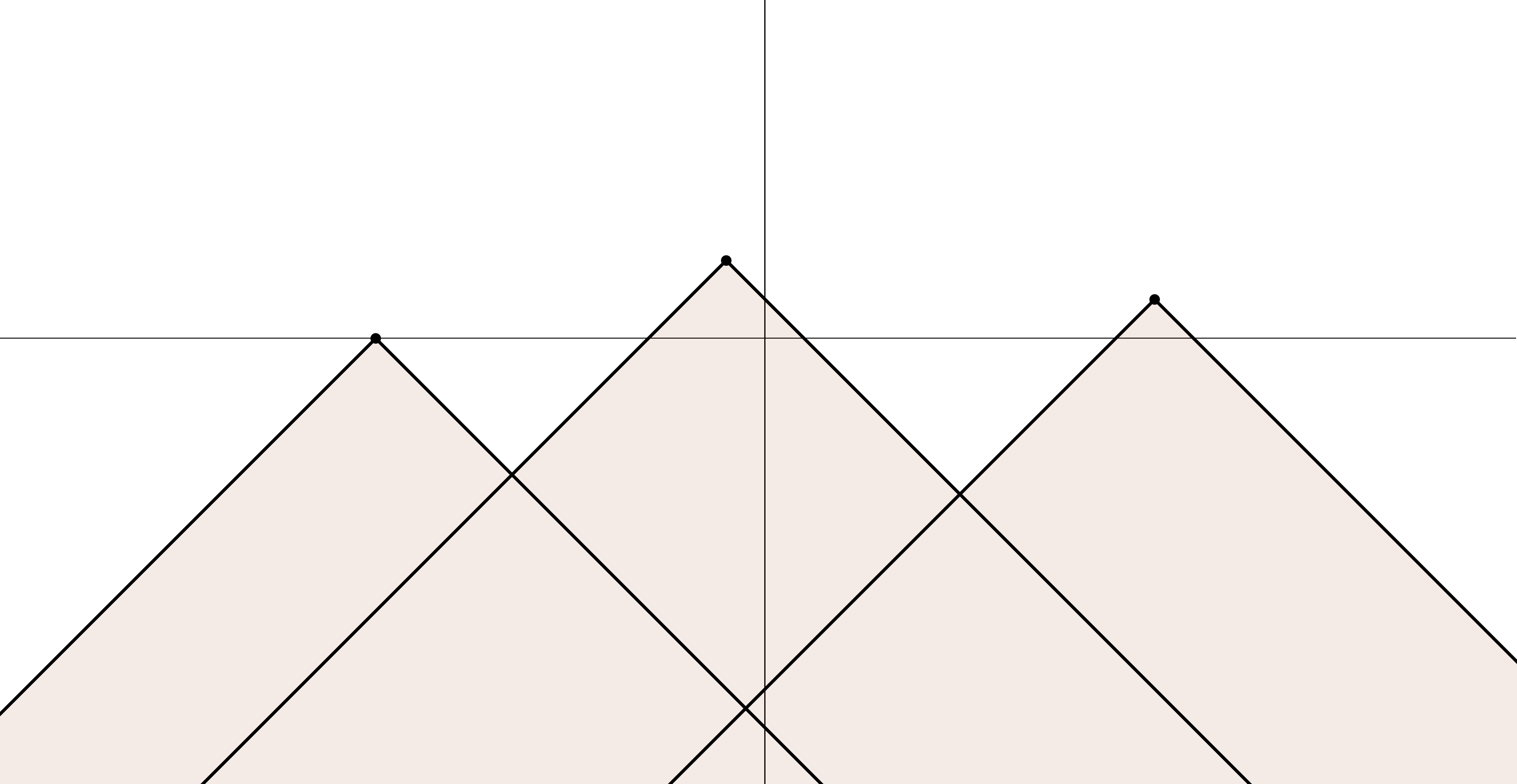} (a)}
\end{minipage}
\begin{minipage}[h]{0.49\linewidth}
\center{\includegraphics[width=1\textwidth]{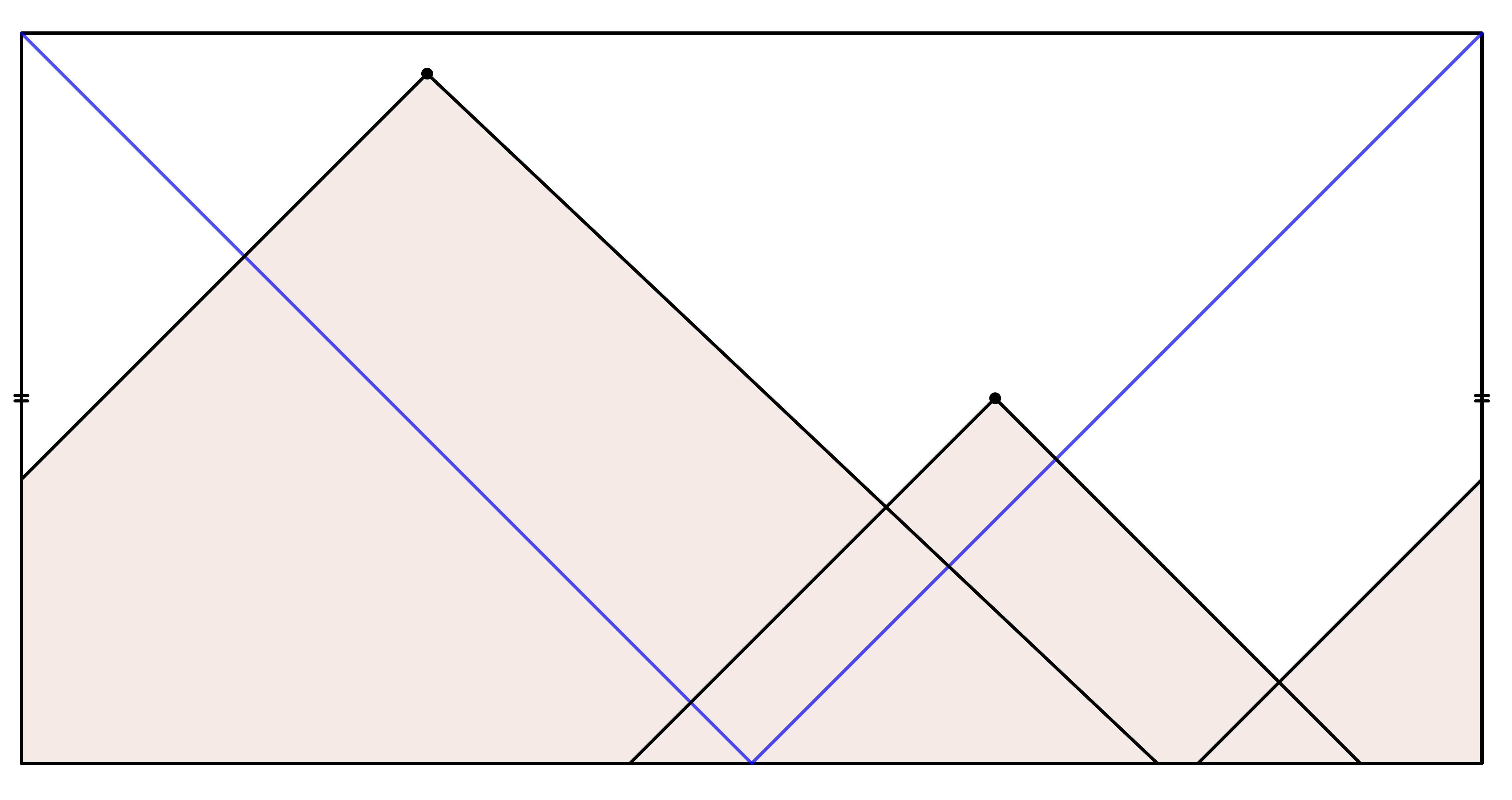} (b)}
\end{minipage}
\caption{The union of past ligh cones of external points of a diagram: (a) in Minkowski space-time; (b) on the Penrose diagram of the two-dimensionsl de Sitter space-time; the blue line shows the boundary between expanding and contracting Poincare patches.}\label{LCones}
\end{figure}

Before we proceed with other charts of the Minkowski and de Sitter space-times, let us discuss one important property of the Schwinger-Keldysh diagrammatic technique. Namely, consider an arbitrary diagram with external points $E_1,\dots,E_m$~--- a contribution to the $m$-point correlator. A property which we will refer to as \emph{causality} is as follows: \emph{The integration over the internal vertices in the diagram can be restricted to the interior of the union of the past light-cones emanating from $E_1,\dots E_m$}, as shown on the Fig. \ref{LCones}.

Note that for regularized propagators one obtains that $F(Y,Y) = \bar{F}(Y,Y)$. Hence, the commutator $\bl [\phi(X),\phi(Y)]\br$ vanishes not only for space-like, but also for light-like separations between $X$ and $Y$ . As in the proof below we will relay only on causal properties of propagators, it is convenient to treat light-like separations similarly to space-like ones. Also from now on by saying <<light-cone>> we will actually mean its interior.

To prove the statement in the first paragraph of this section, let us start from a convenient observation. Consider a vertex $Y$ connected with $n$ points $X_j$, similarly to the subsection \ref{skc}. Note that we do not specify whether $X_j$ are internal or external vertices of the entire diagram (they are external vertices only for the \emph{part} of the diagram under consideration). Thus, we deal with the integral of such a form as (\ref{IK}). Consider its integrand:
\begin{equation}
\hat{F}_n(X_1,\dots,X_n,Y) = \prod_{j=1}^k F(Y,X_j) \prod_{j=k+1}^n \bar{W}(Y,X_j) - \prod_{j=1}^k W(Y,X_j) \prod_{j=k+1}^n\bar{F}(Y,X_j).\label{31}
\end{equation}
First of all, if $Y$ is outside of all the light-cones emanating from $X_j$ (both past and future), this expression vanishes, because the Feynamn and Wightman propagators coincide. In fact, the propagators are real functions for space-like separations between their arguments and the $i\ep$--prescriptions in $F$'s and $W$'s in (\ref{31}) are irrelevant. Hence, the integration over $Y$ in (\ref{IK}) can be restricted to the union of the light-cones with the vertices at $X_j$, $j = \overline{1,n}$.  

Next, assume that $Y$ does not belong to the union of past light-cones.  Let $\mathfrak{S}$ be the set of all such indices that for $j \in \mtf{S}$ we have $X_j^0 \ge Y^0$ (the set obviously can be empty). It means that $Y^0$ cannot be in the future light-cone of $X_j^0$, and by our assumption it cannot be in the past light-cone as well. So for all $j \in \mtf{S}$ the vertex $Y$ is outside of the light-cone of $X_j$. Let us define the following auxiliary objects:
\begin{equation}
\begin{aligned}
\hat{F}_{n,\mtf{S}}^{-} &= \prod_{j=1,\,j\in\mtf{S}}^k F(Y,X_j) \prod_{j=k+1,\,j\in\mtf{S}}^n \bar{W}(Y,X_j),\quad \hat{F}_{n,\mtf{S}}^{+} = \prod_{j=1,\,j\in\mtf{S}}^k W(Y,X_j) \prod_{j=k+1,\,j\in\mtf{S}}^n\bar{F}(Y,X_j),\\
\hat{F}_{n,\bar{\mtf{S}}}^{-} &= \prod_{j=1,\,j\notin\mtf{S}}^k F(Y,X_j) \prod_{j=k+1,\,j\notin\mtf{S}}^n \bar{W}(Y,X_j),\quad \hat{F}_{n,\bar{\mtf{S}}}^{+} = \prod_{j=1,\,j\notin\mtf{S}}^k W(Y,X_j) \prod_{j=k+1,\,j\notin\mtf{S}}^n\bar{F}(Y,X_j).
\end{aligned}
\end{equation}
Then $\hat{F}_n$ from (\ref{31}) can be expressed as follows:
\begin{equation}
\hat{F}_n(X_1,\dots,X_n,Y) = \hat{F}_{n,\mtf{S}}^{-} \hat{F}_{n,\bar{\mtf{S}}}^{-} - \hat{F}_{n,\mtf{S}}^{+} \hat{F}_{n,\bar{\mtf{S}}}^{+}.\label{FnRepr}
\end{equation}
If $j \in \mtf{S}$, the interval between $X_j$ and $Y$ is space-like as we have just assumed and all of the propagators between these points coincide, hence $\hat{F}_{n,\mtf{S}}^{-} = \hat{F}_{n,\mtf{S}}^{+}$. On the other hand, if $j \notin \mtf{S}$, then $Y^0 > X_j^0$, therefore $F(Y,X_j) = W(Y,X_j)$ and $\bar{F}(Y,X_j) = \bar{W}(Y,X_j)$. It means that $\hat{F}_{n,\bar{\mtf{S}}}^{-} = \hat{F}_{n,\bar{\mtf{S}}}^{+}$. Thus, from (\ref{FnRepr}) we obtain that $\hat{F}_n = 0$ and we conclude that \emph{the integrand of the loop correction can be nonzero only if $Y$ belongs to the union of past light-cones of $X_j$ for all $j$}.

Now consider the integrand of the entire graph whose vertices are $V_I$, $I = 1,\dots,N$. The external ones among them we denote as $E_j$, $j = 1,\dots,m$ just as at the beginning of this subsection.  We will use the notation $V_I \prec V_K$ if $V_I$ is in the past light-cone of $V_K$. Note that the relation <<$\prec$>> defines the strict ordering as we consider only the light-cone interiors: $V_K\not\prec V_K$ and if $V_K \prec V_L$ and $V_L \prec V_M$ then obviously $V_K \prec V_M$ due to the structure of light-cones.

Let $V_{J_1}$ be an arbitrary \emph{internal} vertex of the graph. For the integrand of the graph to be nonzero it can be connected only with such an other vertex $V_{J_2}$ that $V_{J_1} \prec V_{J_2}$, as we have shown above in this section. If $V_{J_2}$ is an internal vertex, the chain of relations can be extended further: $V_{J_1} \prec V_{J_2} \prec V_{J_3}$. Due to the properties of <<$\prec$>>, $J_3$ is distinct from $J_1$ and $J_2$. If $V_{J_3}$ is internal, we can continue the extension process: $V_{J_1}\prec V_{J_2} \prec V_{J_3} \prec\dots\prec V_{J_l}$ for some $l$ with pairwise distinct $J_1,\dots,J_l$. Note that all vertices here except for $V_{J_l}$ are necessarily internal. Assume that $V_{J_l}$ is a terminal vertex, e.g. there is no vertex $V_{J_{l+1}}$ such that $V_{J_l} \prec V_{J_{l+1}}$. If $V_{J_{l}}$ is internal it implies that the integrand is zero. Hence, it should be external for the integrand to be non-zero. Thus, e. g. $V_{J_l} = E_m$ for some $m$. 

In all, the above defined causality property of the Schwinger-Keldysh technique is proved. Note that to prove this statement we have used only analytic properties of the propagators, which are related to the presence of the cuts along the time-like separations of their arguments. We did not use in the prove the fact that the space-time is flat. Hence, the statement formulated at the beginning of this section is true for any space-time including the de Sitter one.

\section{Other charts of Minkowski space-time and causality}\label{PPow}
The property of the causality, that we have discussed in the previous subsection, is very useful for the consideration of various patches of Minkowski space-time. To see that let us prove the Poincare invariance of the loop integrals in the case when there is only one theta-function in (\ref{IK}), which restricts to a half of the entire Minkowski space-time rather than to a quarter (quadrant or wedge). In particular, let it be $\theta(Y^0+Y^1)$, i.e. we have the restriction $Y^0 > -Y^1$ and consider half of the entire flat space-time cut off by the light-like surface $Y^0 = -Y^1$. This region is somewhat similar to the expanding Poincare patch of de Sitter space-time which we will consider in the section \ref{EPPSP} below. At the same time the right Rindler wedge is similar to the static patch of the de Sitter space-time. The de Sitter invariance of the loop corrections for the Bunch-Davies state in Poincare patch was discussed in e.g. \cite{Polyakov:2012uc,Akhmedov:2013vka} with the use of the same methods as we apply here. 

The variation of the analog of $I_K$, in which there is only one $\theta(Y^0+Y^1)$ rather than two, is simply $\delta^{(2)}I_K$ from (\ref{Del1Del2}) without $\theta(Y^0)$. Of course, we also remove the condition $X_j^1 > X_j^0$ on the external points as well assuming only that $X_j^1 > -X_j^0$. Let us show that the variation can again be represented as an integral around the cut. First, we can rewrite $d_2(Y,X_j)$ form (\ref{d12}) as follows:
\begin{equation}
d_2(Y,X_j) = (X_j^1+X_j^0) \, \Big[2(Y^0+X_j^0)-(X_j^1+X_j^0)\Big] - (Y^a-X^a_j)^2.
\end{equation}
Let us analyze the connection between the sign of the geodesic distance $d_2$ and signs of $i\ep$-prescriptions in (\ref{Del1Del2}) as in the subsection \ref{skc}. Let us start from the region where $i\ep$-prescriptions in the first and second summands associated with $X_j$ have the same sign, i.e $\sign(Y^0 + X_j^0) = -$. It is easy to see that then $d_2(Y,X_j) < 0$~--- the interval is space-like. Furthermore, it follows that if $d_2(Y,X_j) > 0$, then $\sign (Y^0 + X_j^0) = +$ and the corresponding $i\ep$-prescriptions in the two terms under the integral in (\ref{Del1Del2}) are different from each other. 

In all, we have a situation similar to the one at the end of subsection \ref{skc}: the variation can be expressed as in (\ref{delIK2}) but without $\theta(Y^0)$. Hence the first summand of the variation can be represented as the integral along the real line going below the cut, and the second one~--- going above the cut. Both of these contours can be closed by infinite semicircles, so both summands in (\ref{delIK2}) (without $\theta(Y^0)$) vanish independently.

Similarly one can prove the Poincare invariance of the integral (\ref{IK}) when only $\theta(Y^1-Y^0)$ is present instead of $\theta(Y^1+Y^0)$. As in the subsection \ref{skc} one can use the translations moving the surface $Y^1=0$ to $Y^1 \to -\infty$ to show the coincidence with the perturbation theory in the whole Minkowski space-time.

Having these observations in mind, the Poincare invariance of the loop integrals in the Rindler wedge can also be derived from the causality property: the intersection of the past light-cone of any point in the wedge with the wedge itself is the same as the intersection of the same light-cone with the half of Minkowski space-time, $Y^1>-Y^0$ containing the wedge. So, the result of any calculation in the Schwinger-Keldysh technique in the right wedge coincides with the result of the similar calculation in the half containing the wedge. The analysis of the subsection \ref{skc} is, however, still instructive, as it will help us to consider the situation in the future (or upper) wedge, $X^0 > |X^1|$, of Minkowski space-time. 

In the same manner one can show the Poincare invariance of the loop integrals in the left wedge ($Y^1 < -|Y^0|$). Furthermore, note that a past light-cone of any point in the past wedge $Y^0 < -|Y^1|$ is contained in this wedge, so the the Schwinger-Keldysh perturbation theory there also coincides with one in the entire Minkowski space-time due to the causality property  and, hence, is Poincare-invariant. 

The only remaining part is the future wedge $X^0 > |X^1|$ of Minkowski space-time. This case is more complicated: an intersection of any past light-cone with this wedge cannot be represented as its intersection with any half of the entire space-time. Also it can be shown that unlike any other wedge, intersection of the past light-cone of any point in the future wedge with the wedge itself has a finite volume. It means that due to causality there should be no infrared divergences even for massless theories, unlike the situation in the whole Minkowski space-time. So one needs to consider the future wedge separately.

\subsection{Future or upper wedge}

In the future wedge, $X^0 > |X^1|$, the following local coordinates can be introduced:
\begin{equation}
X^0 = e^\tau \cosh\xi,\quad X^1 = e^\tau \sinh\xi.
\end{equation}
Then, the corresponding induced metric,
\begin{equation}
    ds^2_F = e^{2\tau}(d\tau^2-d\xi^2) - (dX^a)^2, \label{metrF}
\end{equation}
depends on time. Hence, the free Hamiltonian is also time dependent. This is another reason why one has to use the Schwinger-Keldysh technique to do loop calculations in the patch under consideration. The Minkowski (Poincare symmetric) propagator corresponds to a specific choice of the basis of modes in the patch, which was found e.g. in \cite{Akhmedov:2021agm}. {Note that while such a state is not stationary in the future wedge, as the free hamiltonian in that region dependends on time, those modes still diagonalize it in the infinite past and by definition describe the exact quantum evolution of the free theory. In the section \ref{dS} we will see that the modes which do not diagonalize the free hamiltonian in the past lead to severe IR problems.}

We want to check whether loop corrections respect the Poincare isometry or not. Similarly to the subsection \ref{skc} we need to consider construction (a part of a loop integral) of the following form:
\begin{equation}
\begin{aligned}
I_K^F(X_1,\dots,X_n) = \ili d^d Y \theta(Y^0-Y^1)\theta(Y^0+Y^1) &\left[\prod_{j=1}^k F(Y,X_j) \prod_{j=k+1}^n \bar{W}(Y,X_j) - \right.\\
&-\left.\prod_{j=1}^k W(Y,X_j) \prod_{j=k+1}^n\bar{F}(Y,X_j) \right].
\end{aligned}
\end{equation}
The step functions now restrict the integration region to the future wedge, and we should assume that all $X_i$'s, $i=1,\dots, n$, are in the same wedge as well. 

\begin{figure}[ht!]
\centering\includegraphics[width=0.75\textwidth]{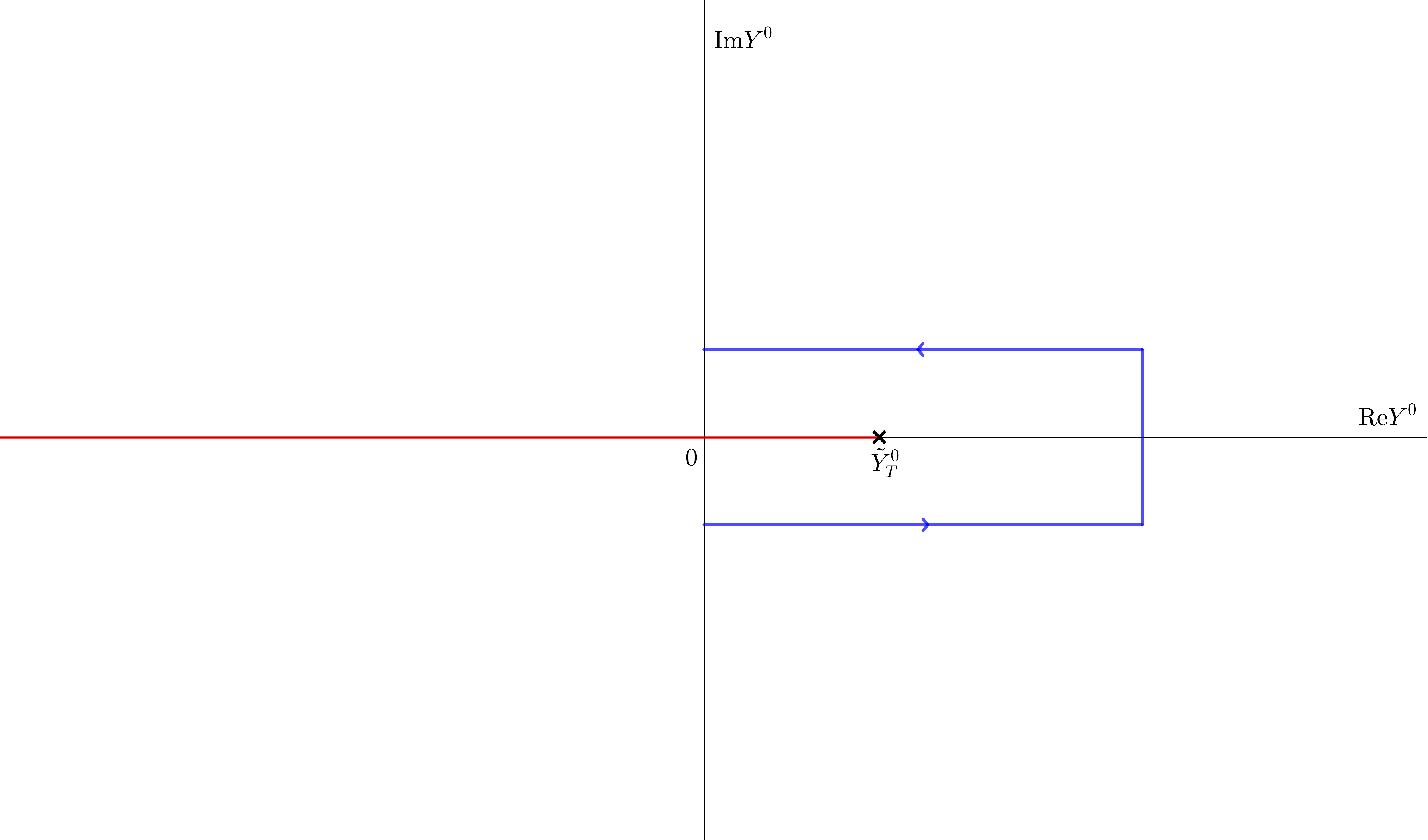}
\caption{The integration contour for $\delta^{(1)}I_K^F$ (blue), the branch cut is shown in red.} \label{I1ContF}
\end{figure} 

The variation of $I_K^F$ under the time translation $Y^0 \to Y^0 + \ep$ can be represented as $\delta_\ep I_K^F = \ep(\delta^{(1)} I_K^F + \delta^{(2)} I_K^F)$ similarly to (\ref{delIK}). Consider, for instance, $\delta^{(1)}I_k^F$. It coincides with $\delta^{(1)}I_K$ from (\ref{Del1Del2}):
\begin{equation}
\begin{aligned}
\delta^{(1)} I_K^F &= \ili d^{d-1} Y \theta(Y^0)\left\{ \prod_{j=1}^k\mathfrak{F}\Big[d_1(X_j,Y) - i\ep\Big] \prod_{j=k+1}^n \mathfrak{F}\Big[d_1(X_j,Y) + i\ep \sign(Y^0 - X_j^0)\Big]-\right.\\
&-\left. \prod_{j=1}^k\mathfrak{F}\Big[d_1(X_j,Y) - i\ep \sign(Y^0-X^0_j)\Big] \prod_{j=k+1}^n \mathfrak{F}\Big[d_1(X_j,Y) + i\ep\Big]\right\}.
\end{aligned}\label{del1IKF}
\end{equation}
Note that $d_1(X_j,Y)$ from (\ref{d12}) can be rewritten as follows:
\begin{equation}
d_1(Y,X_j) = (X_j^1-X_j^0)\, \Big[2(Y^0- X_j^0) + X_j^0 - X_j^1\Big] - (Y^a-X_j^a)^2.\label{d1Re}
\end{equation}
Let us repeat the analysis of $i\ep$-prescriptions once again. As before, we will start with the region where $i\ep$-prescriptions of two terms of the integrand of (\ref{del1IKF}) have the same signs, e.g. $\sign (Y^0 - X_j^0) = +$. It follows from the definition of the upper wedge and from (\ref{d1Re}) that $d_1(Y,X_j) < 0$. So just as in the case considered above we can write:
\begin{equation}
\begin{aligned}
\delta^{(1)} I_K^F &= \ili d^{d-1} Y \theta(Y^0)\left\{ \prod_{j=1}^n\mathfrak{F}\Big[d_1(X_j,Y) - i\ep\Big] - \prod_{j=1}^n\mathfrak{F}\Big[d_1(X_j,Y) + i\ep\Big]\right\}.
\end{aligned}
\end{equation}
However, the key difference from the cases considered above in this section and in the subsection \ref{skc} is a position of the cut. From (\ref{d1Re}) it is clear that $d_1(Y,X_j) > 0$ if $Y^0$ is negative and $|Y^0| \gg 1$, meaning that the cut is going along $(-\infty,\tilde{Y}_t^0)$. Here $\tilde{Y}_t^0$ is the highest value of $Y^0$ such that $d_1(Y,X_j) = 0$ for some $j$. Note that $d_1(Y,X_j) = (X_j^0)^2 - (X_j^1)^2 - (Y^2-X_j^a)^2$ if $Y^0 = 0$. It can be positive for some values of $X_j^\mu$ and $Y^a$ and in such cases $\tilde{Y}_t^0 > 0$. This situation is shown on the Fig. \ref{I1ContF}. Note that the integration goes around only a part of the cut $(0,\tilde{Y}_t^0)$. Hence it cannot be properly closed by an infinite circle and is not zero in generic situations.

The same way one can show that $\delta^{(2)} I_K^F \neq 0$ in generic situations. So the Schwinger-Keldysh technique in the future wedge does not provide the isometry invariant expressions.
The reason why we encounter the differences between loop calculations in the entire Minkowski space-time and in the future wedge, although we use the same tree-level propagators, is as follows: Cauchy surfaces in the entire Minkowski space-time and in the future wedge have different geometry. That results in different infrared effects in these different charts, as these effects are sensitive to the boundary and initial conditions. We will encounter a similar situation in various patches in de Sitter space-time below.

Let us demonstrate the violation of the isometry-invariance in the future wedge on a simple example. Note that if the correlation functions in the theory are Poincare-invariant, they should coincide with those in the theory in the entire Minkowski space-time as we can move the surface $Y^0 = 0$ to $-\infty$ via time translations (part of the Poincare group). So it is enough to show that the result of some calculation is different in entire Minkowski space-time and in the future wedge. Consider the massless scalar field theory in $d=4$. Its Wightman function is as follows:
\begin{equation}
W(X,Y) = \dfrac{1}{(X-Y)^2 - i\ep \sign(X^0-Y^0)}.
\end{equation}
We can add the mass term $\frac{m^2 \phi^2}{2}$ to the lagrangian as a perturbation and consider the first-order correction $F^{(1)}(0,X)$ to the Feynman propagator $F(0,X)$, where $X^\mu = (t,0,0,0)$. Of course the point $0$ is on the boundary of the right wedge, but it can be considered as a limit of some sequence of points inside the wedge. In the entire Minkowski space-time we can use the standard Feynman perturbation theory to express the correction:
\begin{equation}
F^{(1)}_M(0,X) = -im^2 \ili d^4 Y \dfrac{1}{(Y^2-i\ep)((Y-X)^2-i\ep)}.\label{FM}
\end{equation}
The Schwinger-Keldysh technique in the entire Minkowski space-time for the static state that we consider here will give the same result.

In the future wedge we have to use the Schwinger-Keldysh technique, as was explained above. First, note that for the points under consideration $F(Y,0) = W(Y,0)$ as in the future wedge $Y^0 > 0$. Then, note that
\begin{equation}
F(Y,X) - W(Y,X) = \dfrac{1}{(Y-X)^2-i\ep} - \dfrac{1}{(Y-X)^2 - i\ep \sign(Y^0-t)} = 2i\pi\theta(t-Y^0) \delta[(Y-X)^2].
\end{equation}
As a result, in the future wedge the correction to the propagator for the two points under consideration is as follows:
\begin{equation}
F^{(1)}_F(0,X) = 2\pi m^2 \ili_{|Y^1|<Y^0<t} d^4Y \dfrac{\delta[(Y-X)^2]}{(Y^2-i\ep)}.\label{FF}
\end{equation}

The integrals (\ref{FM}) and (\ref{FF}) are calculated in the appendix \ref{propcalc}. The results are:
\begin{equation}
F^{(1)}_M(0,X) = -\pi^2 m^2 \log \dfrac{\Lambda^2}{-t^2},\quad F^{(1)}_F(0,X) = i\pi^3m^2, \label{FMFF}
\end{equation}
where $\Lambda$ is an IR cutoff. The results are clearly different (but their imaginary parts, however, do coincide). In the entire Minkowski space-time for the massless theory there is an IR divergence which becomes stronger with the increase of the order of perturbation theory. So one has to sum the whole series to obtain the correct result, which is divergence-free and non-analytic in $m^2$. 

On the other hand the intersection of the future wedge with past light-cone of an arbitrary point in it has a finite volume as we pointed out above. Hence, due to the causality property even the massless theory in this chart is free from the IR divergences. This is definitely the case for $F_F^{(1)}$, as one can see from the result. Hence, the theory in the future wedge cannot be Poincare-invariant.

It is also worth stressing that the consideration above also works in the case of massive theory with mass $\mu$ such that $t \ll \frac{1}{\mu}$, as $\frac{1}{\mu}$ is a natural cutoff scale in this case. The characteristic size of the past light-cone of $X$ is $\sim t$, so the theory is effectively massless and the expression for $F_F^{(1)}(0,X)$ from (\ref{FMFF}) is still valid. On the other hand, we can consider $F_F^{(1)}(X_1,X_2)$, where $X_1^\mu = (T,0,0,0)$ and $X_2^\mu = (T+t,0,0,0)$ with $T \gg \frac{1}{\mu}$. The principal contribution to the integral in the perturbation theory comes from the region around $X_1$ and $X_2$ of the size $\sim\frac{1}{\mu}$, which is entirely contained in the intersection of the past light-cone of $X_2$ with the future wedge. Hence one can use the Minkowski space-time calculation and the first-order correction is given by $F_M^{(1)}$ from (\ref{FMFF}) where $\Lambda$ is replaced with $\frac{1}{\mu}$. Now we explicitly see the violation of the isometry. In particular, it follows that an adiabatic change of the mass in the past moves the system from the isometry-invariant state to some other state. Similar effects in the case of the global de Sitter space-time were observed in \cite{Polyakov:2022dpa}.

Another interesting consideration is that one can modify the Schwinger-Keldysh technique to make the corrections under consideration to be isometry-invariant. Namely, the arguments about analytic continuation from the subsection \ref{EuCont} work in the whole Minkowski space-time as well, but one can move the horizontal parts of the contour $C$ from the Fig. \ref{AnCont} to positive and negative imaginary infinities. This is possible as there is no periodicity in imaginary time and therefore no additional cuts from the Fig. \ref{RinCuts} and we simply obtain the perturbation theory in the euclidean $\mathbb{R}^d$. However, while on the Fig. \ref{EuCont} the euclidean contour is closed in the left half plane, one can close it in the right half plane as well. This allows us to redefine the calculations in such a way that future and past infinities are effectively interchanged. The contour now starts at future infinity above all cuts and returns to it below them.

Thus, in this modification of the Schwinger-Keldysh technique one exchanges the future and past infinities in the calculations, effectively mapping the calculation from the future wedge to the one in the past wedge. Namely the <<$-$>> and <<$+$>> parts of the contour change their directions (<<$-$>> part goes from future to past and <<$+$>> part from past to future). It means that we should assign $-1$ multipliers to <<$+$>> vertices instead of <<$-$>> ones. Besides that, $G_{--}(Y,X_j)$ is now below the cut if $Y^0 < X_j^0$ and above it if $Y^0 >X_j^0$~--- the <<$-$>> part is above cuts in the future as we discussed. Hence, we need to conjugate it and define $G_{--}(Y,X_j) = \bar{F}(Y,X_j)$. In the same way $G_{++}(Y,X_j) = F(Y,X_j)$. Functions $G_{-+}(Y,X_j)$ and $G_{+-}(Y,X_j)$ are still above and below the cut respectively, so they are unchanged.
 
Using the argument similar to the one in the section \ref{Caus} one can show that the integration regions for internal vertices can be restricted to the union of \emph{future} light-cones emanating from the external points. Hence, this technique in the upper wedge is the same as in the entire Minkowski space-time (with the reversed time) and is Poincare-invariant as a result. However, unlike the case with the usual Schwinger-Keldysh formalism it is questionable whether such a technique has some sensible physical interpretation.

{It is also instructive to note that one might choose different modes than in \cite{Akhmedov:2021agm}. However, as we will discuss below in the section \ref{dS}, another choice of the basis of modes might lead to additional UV problems or the corresponding Fock space ground state will violate the isometry. Besides that, the proof of causality property above is based on the fact that the commutator of the field operators vanishes outside the light cone. The commutator is the c-numbers in the free theory, so it does not depend on the choice of the basis of modes. Hence, the causality always holds, so the argument regarding the finite volume of the past light cone always holds and we will never get IR divergences in the future wedge. Therefore, any choice of modes cannot lead to the same one-loop result as in the entire Minkowski space-time, which has IR divergences in the massless theory.}

Finally, one can consider the following euclidean version of the metric (\ref{metrF}):
\begin{equation}
    ds^2_{F,E} = e^{-2i\kappa}(d\kappa^2+d\xi^2) + (dX^a)^2.\label{FEmetr}
\end{equation}
It is $2\pi$-periodic in $\kappa$ just as the usual static Rindler metric, but also it is complex valued. The question whether a space with a complex euclidean metric admits a sensible quantum field theory in static sense was addressed in \cite{Kontsevich:2021dmb,Witten:2021nzp}. The criterion formulated by Kontsevich and Segal is as follows: if the metric has a form
\begin{equation}
    ds^2 = \sum_{i=1}^d \lambda_i dx_i^2,
\end{equation}
then the inequality $\sum_{i=1}^d |\text{Arg}\,\lambda_i| < \pi$ where arguments are defined in $(-\pi,\pi)$ should be satisfied in each point of the manifold. This is clearly not the case for (\ref{FEmetr}) if e.g. $\kappa = \frac{\pi}{2}$, so the QFT in the upper wedge does not have a euclidean version.

\section{Charts of the de Sitter space-time}\label{dS}
In this section we will consider various patches of the $d$-dimensional de Sitter (dS) space-time. Our presentation in the expanding Poincare patch in many respects repeats that of \cite{Hollands:2014eia,Higuchi:2010xt}. 

dS space-time is defined as a hypersurface embedded in the $(d+1)$-dimensional Minkowski space-time with coordinates $X^I$, $I=0,\dots,d$ whose metric has signature $(+,-,\dots,-)$. The equation defining the hypersurface of unit curvature is as follows:
\begin{equation}
    X_I X^I = -1.
\end{equation}
It admits the following global coordinates:
\begin{equation}
X^0 = \sinh t,\quad X^\sigma = \psi^\sigma \cosh t,~\sigma = 1,\dots,d \, ,
\end{equation}
where $\psi^\sigma$ are components of a vector on a $(d-1)$-dimensional sphere: $(\psi^\sigma)^2 = 1$. The corresponding metric is as follows:
\begin{equation}
    ds^2 = dt^2 - \cosh^2 t\,d\Omega_{d-1}^2,\label{GdSmetr}
\end{equation}
here $d \Omega_{d-1}^2$ is a metric on a $(d-1)$-dimensional unit sphere. It is also convenient to introduce the geodesic parameter $\zeta(X,Y) = -X_IY^I$ which is invariant with respect to the dS space-time isometries~--- the Lorentz group of the embedding Minkowski space-time. The sign is chosen in such a way that $\zeta(X,Y) = \frac{(X-Y)^2}{2}+1$. So the interval between $X$ and $Y$ is space-like if $\zeta < 1$ and time-like if $\zeta > 1$. In global coordinates we have that
\begin{equation}
\zeta(X,X') = \cosh(t-t') - (1-\psi^\sigma\psi'^\sigma)\cosh t\,\cosh t'.
\end{equation}
Hence, the standard $i\ep$-prescription $t-t'-i\ep$ for the Wightman function translates into $\zeta(X,X') - i\ep\sign(t-t')$, which is similar to $i\ep$-prescription in the Minkowski space-time of the interval squared. Aslo note that $\sign(t-t') = \sign(X^0-X'^0)$.

As before, we consider the scalar QFT in different patches of dS space-time and check whether it is isometry-invariant on the loop level if one chooses invariant states (and, correspondingly, propagators) at tree-level. The isometries in this case are given by the connected component of unity of the Lorentz group $SO(d-1,1)$ of the embedding Minkowski space-time. As the metric (\ref{GdSmetr}) is time-dependent, there is no canonical choice of positive energy modes. However, there is one particular choice that corresponds to a state in which isometry-invariant correlators are related to similar correlators in the euclidean theory on $S^d$ via analytic continuation. In such a state correlation functions obey the Hadamard conditions, i.e. behave as in flat space-time for short separations. 

This maximally-analytic state is usually referred to as the Bunch-Davies (BD) vacuum \cite{Bunch:1978yq}. To simplify notations we will denote the corresponding Wightman function by $W(X,X')$ as before in the case of Mikowski space-time. It is as follows \cite{Mottola:1984ar,Akhmedov:2013vka,Allen:1985ux,Bros:1995js,Akhmedov:2019esv}:
\begin{equation}
W(X,X') = \dfrac{\Gamma\lb\frac{d-1}{2}+i\mu\rb \Gamma\lb\frac{d-1}{2}-i\mu\rb}{(4\pi)^{\frac{d}{2}}\Gamma\lb \frac{d}{2}\rb}\,_2F_1\lb\frac{d-1}{2}+i\mu,\frac{d-1}{2}-i\mu,\frac{d}{2};\frac{1+\zeta-i\ep\sign(t-t')}{2} \rb,\label{BDW}
\end{equation}
where
\begin{equation}
    \mu^2 = m^2-\dfrac{(d-1)^2}{4}
\end{equation}
and $\,_2F_1(a,b,c;x)$ is a standard hypergeometric function with the branch cut along $(1,+\infty)$. For more details about analytic properties of this propagator see e.g. \cite{Bros:1994dn,Bros:1995js,Bros:1998ik}. Here $m$ is the mass of the scalar field measured in units of dS curvature, which we set to be one. Such a Wightman function is clearly isometry-invariant. As $\zeta(X,X') = \frac{(X-X')^2}{2}+1$, the expression in terms of analytic function of the embedding Minkowski space-time interval (\ref{Wiep}) can be used as well. 

This is not the only choice of modes which leads to the isometry-invariant propagators in the dS space-time. In fact, there is a continuous family of so-called alpha states respecting the symmetry   \cite{Mottola:1984ar,Allen:1985ux}. The corresponding Wightman functions can be expressed in terms of the BD one (\ref{BDW}) \cite{Epstein:2014jaa}:
\begin{equation}
W^{(\alpha)}(X,X') = \cosh^2\alpha\,W(X,X') + \sinh^2\alpha\,W(X',X) + \sinh\alpha\cosh\alpha\,\Big[W(X,-X') + W(-X,X')\Big].\label{Walpha}
\end{equation}
In the case $\alpha = 0$ it coincides with the BD Wightman function, but for generic values of $\alpha$ it has different analytic properties and does not obey the Hadamard conditions. For instance, unlike $W(X,X')$ it does not admit the representation of the form (\ref{Wiep}): $W(X,X')$ and $W(X',X)$ have opposite $i\ep$-prescriptions. As we will discuss in the subsection \ref{EPPSP}, this fact has drastic consequences in perturbation theory.

In loop integrals we need to use the dS-invariant volume form, $d\text{Vol}_{dS}$, defined by the metric (\ref{GdSmetr}). It is, however, more convenient for our purposes to write all integrals in terms of the embedding Minkowski space-time coordinates $Y^I$ inserting $\delta(Y^2+1)$ into the integrand. More precisely, we can use the following expression for the volume form:
\begin{equation}
    d\text{Vol}_{dS} = 2\delta(Y^2+1)d^{d+1}Y.\label{dSMeas}
\end{equation}
In these notations all of the integrals are manifestly isometry-invarinat if we integrate over the whole dS space-time, but contain IR divergences, as we will discuss below. Besides that, the Schwinger-Keldysh integrals have the form similar to (\ref{IK}) with the measure defined above (the step functions should of course be used only if we wish to restrict the integration domain to some region of the entire space-time). 

\subsection{Loop corrections in the Expanding Poincare and static patches}\label{EPPSP}
The expanding Poincare patch (EPP) is a half of entire dS space-time defined e.g. by the condition $X^0 > -X^d$. It admits the following local coordinates:
\begin{equation}
X^0 = \sinh\tau + \dfrac{(x^i)^2}{2}e^\tau,\quad X^i= x^i e^\tau,\quad X^d = \cosh\tau - \dfrac{(x^i)^2}{2}e^\tau,~ i = 1,\dots,d-1,\label{EPPcoord}
\end{equation}
the metric for $x^i$ is $\delta_{ij}$. The induced metric on EPP from the ambient one is as follows:
\begin{equation}
    ds^2 = d\tau^2 - e^{2\tau} (dx^i)^2.
\end{equation}
Usually the conformal time $\eta = e^{-\tau}$ is used instead of of the proper one $\tau$ as then the metric becomes conformally-flat:
\begin{equation}
ds^2 = \dfrac{d\eta^2 - (dx^i)^2}{\eta^2}.\label{ConfMet}
\end{equation}
For explicit expressions of modes corresponding to BD and alpha states in the EPP see e.g. \cite{Akhmedov:2019esv}.

We want to check whether the Schwiger-Keldysh technique for the BD state in the EPP is isometry-invariant. In the section \ref{PPow} we have considered the half of Minkowski space-time defined by the similar condition $X^0 > -X^1$. As we noted above, the BD propagators have the same analytic representations (\ref{Wiep}) as in the Minkowski space-time, so the only modification to the analysis from the section \ref{PPow} is the use of the measure (\ref{dSMeas}) instead of $d^{d+1}X$. We also need to consider the Lorentz transformations of the ambient space-time instead of translations, but the variation of $\theta(Y^0+Y^d)$ defining the EPP is still proportional to $\delta(Y^0+Y^d)$. Note that this $\delta$-function turns $\delta(Y^2+1)$ into $\delta[1-(Y^i)^2]$, which does not affect the $Y^0$ integration domain and analytic properties of the integrand with respect to $Y^0$. Therefore one can perform the same contour manipulations as in the sections \ref{skc} and \ref{PPow} and find that the Schwinger-Keldysh perturbation theory in EPP is isometry-invariant.

Another interesting region is the static patch (SP) $X^d > |X^0|$~--- it is defined in a similar way as the right Rindler wedge of the Minkowski space-time considered in the section \ref{RinWed}. The SP is half of EPP and, hence, is quarter of the whole dS space-time. The local coordinates in the patch are as follows:
\begin{equation}
X^0 = \sin\theta \sinh t,\quad X^i = \cos\theta\,\psi^i,\quad X^d = \sin\theta\cosh t,~\theta \in [0,\pi/2],~(\psi^i)^2 = 1.
\end{equation}
The main advantage of these coordinates is that the corresponding metric
\begin{equation}
    ds^2 = \sin^2\theta\, dt^2 - d\theta^2 - \cos^2\theta\,d\Omega_{d-2}^2\label{SPmetr}
\end{equation}
is time-independent, so we can define positive energy modes and thermal states. Besides that, the change $t = -i\kappa$ with $\kappa \in (-\pi,\pi)$ transforms (\ref{SPmetr}) into the metric on euclidean $S^d$ of unit radius. It is somewhat similar to the Rindler space being analytic continuation of the euclidean space $\mathbb{R}^d$.

It turns out that the BD propagator corresponds to the thermal state in the static patch with the inverse temperature $\beta=2\pi$ \cite{Gibbons:1977mu,Bros:1995js,Akhmedov:2020qxd,Akhmedov:2021cwh}~--- another similarity with the Rindler space. So as the metric is time-independent one can use exactly the same method as in the subsection \ref{EuCont} to show that the Schwinger-Keldysh technique for BD propagators in the SP is the analytic continuation of the euclidean perturbation theory on $S^d$. Hence, the diagrammatic technique for BD state in the SP is manifestly isometry-invariant. 

Moreover, light-cones in dS space-time are just intersections of ambient space-time light-cones with the dS manifold itself. It means that the causality argument, regarding the wedges $Y^1>|Y^0|$ and $Y^1>-Y^0$ of the Minkowski space-time from the section \ref{PPow}, also works in the dS space-time. Namely, the past light-cone of a point in SP coincides with the past light-cone of the same point in EPP, so the Schwinger-Keldysh perturbation theories for BD states in these charts of dS space-time \emph{exactly coincide}. In particular, we can use the analytic continuation to $S^d$ for the BD state in the EPP as well. Some concrete computations of loop corrections in euclidean signature with subsequent analytic continuation can be found in \cite{Marolf:2010zp, Marolf:2010nz,Marolf:2012kh}.

Finally, we can consider the Schwinger-Keldysh technique for the alpha states\footnote{Here we ignore the fact that for alpha states the leading UV renormalization in dS space-time is different form the one in Minkowski space-time. This means that alpha states are not physically meaningful in the bare UV theory, because otherwise by just measuring the running of coupling constants in local experiments one can sense that one lives in the huge dS space-time.} with $\alpha \neq 0$. But as we discussed, the propagator (\ref{Walpha}) cannot be represented as the boundary value of an analytic function of $\zeta$, so the arguments with contour deformations from the sections \ref{skc} and \ref{PPow} do not work. In particular, let us look at $\delta^{(2)}I_K$ from (\ref{Del1Del2}). In the subsection \ref{skc} we argued that it can be represented as the inegral around the cut as the $i\ep$ prescriptions of all propagators in each summand were the same (but different for each of the two summands). On the other hand, if we use the propagator (\ref{Walpha}) instead of the BD one, additional terms appear where some of the functions $\mtf{F}$ have reversed $i\ep$-prescriptions. Hence, e.g. in the first summand some of the propagators are above the cut and not below it. It means that we cannot close the contour at all and $\delta^{(2)}I_K$ is not zero. 

Besides that, it was crucial for our argument that all of the external points belong to the same patch, but (\ref{Walpha}) also contains terms of the form $W(X,-X')$ and $-X'$ is in a different patch. Hence, the perturbation theory for alpha states in patches of the dS space-time is not isometry-invariant and an adiabatic change of the mass term or turning on the interaction moves the system away from the invariant state.

Another problem of alpha states is the possible appearance of terms like $W(X,X')W(X',X)$ as integrands in the perturbation theory. As it was shown in \cite{Akhmedov:2020jsi}, such terms can lead to sever non-local UV divergences in the effective action which span along the whole light-cone of the point $Y$. This is another revelation of the fact that alpha states have problems with the UV renormalization.

\subsection{One-loop correction in the contracting Poincare patch}\label{1-loopP}
The contracting Poincare patch (CPP) is just the time reversal of EPP and its complement within the global dS space-time: it is defined by the condition that $X^0 < -X^d$. The local coordinates are very similar to (\ref{EPPcoord}):
\begin{equation}
X^0 = \sinh\tau - \dfrac{(x^i)^2}{2}e^{-\tau},\quad X^i= x^i e^\tau,\quad X^d = -\cosh\tau + \dfrac{(x^i)^2}{2}e^{-\tau},~ i = 1,\dots,d-1.\label{CPPcoord}
\end{equation}
The conformal time is now $\eta = e^{\tau}$, so it flows from $\eta = 0$ to $\eta = +\infty$ (in the reverse direction wrt the EPP time), the metric in the conformal time is the same as (\ref{ConfMet}).

In principle, the Schwinger-Keldysh technique for the BD state in the CPP should be manifestly isometry-invariant: due to causality the integration domains in vertices can be extended to the whole dS space-time. However as it was demonstrated in \cite{Akhmedov:2008pu,Akhmedov:2011pj,Polyakov:2012uc,Akhmedov:2013vka} infrared divergences at $\eta \to 0$ appear even if we use the BD propagator for massive theories. 

In this subsection we will show that even the first loop correction in the CPP breaks the dS isometry due to explicit IR divergences. {As we will see, they are present for all values of the mass. Note that if the mass is sufficiently small, IR divergences of a different kind appear even in EPP. They are, however, similar to the usual IR divergences in massless theories in Minkowski space and admit a systematic treatment \cite{Serreau:2013psa, LopezNacir:2016gzi, Gautier:2015pca, Beneke:2012kn,Hollands:2011we}.} To regularize the standard UV divergences just as before we assume to work with Pauli-Villars regularization defined in the appendix \ref{PVreg}, which respects the analytic properties of the propagators.


The version of the Schwinger-Keldysh perturbation theory with <<$+$>> and <<$-$>> vertices is rather inconvenient for practical computations, so let us perform the Keldysh rotation as in \cite{Kamenev,Akhmedov:2013vka}:
\begin{equation}
\phi_{\text{cl}} = \dfrac{\phi_-+\phi_+}{2},\quad \phi_{\text{q}} = \phi_--\phi_+.\label{KeldRot}
\end{equation}
After such a rotation the one loop correction to the Keldysh propagator in the $\phi^3$ theory is as follows (see  \cite{Akhmedov:2012dn,Akhmedov:2019cfd,Akhmedov:2013vka}, for the details):  

\begin{equation}
\begin{split}
    D_1^K (p | \eta_1, \eta_2) = \lambda^2 \int \frac{d^{d-1}\vec{q_1} d^{d-1}\vec{q_2}}{(2\pi)^{2(d-1)}} \int_{\eta_0}^{\infty} \frac{d\eta_3 d\eta_4}{(\eta_3\eta_4)^d} \delta(\vec{p} + \vec{q_1} + \vec{q_2}) \\
    \Big[D_0^R(p | \eta_1, \eta_3) D_0^K(q_1 | \eta_3, \eta_4) D_0^K(q_2 | \eta_3, \eta_4) D_0^A(p | \eta_4, \eta_2) + \\
    + 2 D_0^R(p | \eta_1, \eta_3) D_0^R(q_1 | \eta_3, \eta_4) D_0^K(q_2 | \eta_3, \eta_4) D_0^K(p | \eta_4, \eta_2) +  \\
    + 2 D_0^K(p | \eta_1, \eta_3) D_0^K(q_1 | \eta_3, \eta_4) D_0^A(q_2 | \eta_3, \eta_4) D_0^A(p | \eta_4, \eta_2) -  \\
    - \frac{1}{4} D_0^R(p | \eta_1, \eta_3) D_0^R(q_1 | \eta_3, \eta_4) D_0^R(q_2 | \eta_3, \eta_4) D_0^A(p | \eta_4, \eta_2) - \\
    - \frac{1}{4} D_0^R(p | \eta_1, \eta_3) D_0^A(q_1 | \eta_3, \eta_4) D_0^A(q_2 | \eta_3, \eta_4) D_0^A(p | \eta_4, \eta_2)\Big],
\end{split}
\end{equation}
where $\eta_0$ is the moment after which the self-interactions are adiabatically turned on, $0 < \eta_0 < \eta_1, \eta_2 < +\infty$, i.e. $\eta_0$ is the position of the initial Cauchy surface. In the above expression we use spatially Fourier transformed Keldysh, retarded and advanced propagators, which have the following form:
\begin{align}
\begin{array}{l}
    D_0^K (p | \eta_1, \eta_2) = (\eta_1 \eta_2)^{\frac{d-1}{2}} \text{Re}\Big[h(p\eta_1) \bar{h}(p\eta_2)\Big], \\
    D_0^R (p | \eta_1, \eta_2) = 2 \theta(\eta_1 - \eta_2) (\eta_1 \eta_2)^{\frac{d-1}{2}} \text{Im}\Big[h(p\eta_1) \bar{h}(p\eta_2)\Big], \\
    D_0^A (p | \eta_1, \eta_2) = - 2 \theta(\eta_2 - \eta_1) (\eta_1 \eta_2)^{\frac{d-1}{2}} \text{Im}\Big[h(p\eta_1) \bar{h}(p\eta_2)\Big],\label{PropDef}
\end{array}
\end{align}
where $h(p\eta)=\frac{\sqrt{\pi}}{2 } e^{\frac{\pi \mu}{2}}H_{i\mu}^{(2)}(p\eta)$ is the Hankell function, which defines the BD modes.

In the limit $\eta_0\to 0$ and $\sqrt{\eta_1 \eta_2} \equiv \eta \gg \eta_1/\eta_2$ the leading one-loop contribution to $D_1^K (p | \eta_1, \eta_2)$ can be expressed as:

\begin{equation}
\begin{split}
    D_1^K (p | \eta_1, \eta_2) \approx (\eta_1 \eta_2)^{\frac{d-1}{2}} \Big[h(p\eta_1)\bar{h}(p\eta_2) n_p(\eta) + h(p\eta_1)h(p\eta_2) \kappa_p(\eta) + \text{c.c.} \Big],
\end{split}\label{1-loop}
\end{equation}
where c.c. means complex conjugated terms, meanwhile:
\begin{equation}
\begin{split}
\label{n}
n_p(\eta) \approx \frac{8\lambda^2}{(2\pi)^{2(d-1)}} \int d^{d-1}\vec{q_1} d^{d-1}\vec{q_2} \int_{\eta_0}^{\eta} d\eta_3 d\eta_4 (\eta_3 \eta_4)^{\frac{d-3}{2}} \delta(\vec{p} + \vec{q_1} + \vec{q_2}) \times \\
\times \bar{h}(p\eta_3)h(p\eta_4)\bar{h}(q_1\eta_3)h(q_1\eta_4)\bar{h}(q_2\eta_3)h(q_2\eta_4),
\end{split}
\end{equation}
and

\begin{equation}
\begin{split}
\label{kappa}
\kappa_p(\eta) \approx -\frac{16\lambda^2}{(2\pi)^{2(d-1)}} \int d^{d-1}\vec{q_1} d^{d-1}\vec{q_2} \int_{\eta_0}^{\eta} d\eta_3 \int_{\eta_0}^{\eta_3}d\eta_4 (\eta_3 \eta_4)^{\frac{d-3}{2}} \delta(\vec{p} + \vec{q_1} + \vec{q_2}) \times \\
\times \bar{h}(p\eta_3)\bar{h}(p\eta_4)\bar{h}(q_1\eta_3)h(q_1\eta_4)\bar{h}(q_2\eta_3)h(q_2\eta_4).
\end{split}
\end{equation}
For the scalar field with such a mass that $m > (d-1)/2$ in the limit $p\eta_{3, 4} \ll \mu$  we can use the following asymptotic expansion of the Hankel function:
\begin{equation}
    h(p\eta_{3, 4}) \approx A_{+}(p\eta_{3,4})^{i\mu} + A_{-}(p\eta_{3,4})^{-i\mu},\label{HankExp}
\end{equation}
where $A_\pm$ are some known complex constants whose explicit form is not important for our consideration at this point. We will see now that even for the heavy fields in the CPP and global dS one encounters infrared divergences.

Then the largest contribution to $n_p(\eta)$ and $\kappa_p(\eta)$ in the limit under consideration comes from the region where $q_{1, 2} \gg p$, because for $q_{1, 2} \lesssim p$ the integrals are finite. Hence, we neglect $\vec{p}$ in $\delta$-functions under the integrals on the RHS of (\ref{n}) and (\ref{kappa}). Then we make the following change of variables under the integrals: $u = \eta_3 \eta_4$, $v = \frac{\eta_3}{\eta_4}$, $ \vec{x_1} = \eta_3 \vec{q_1}$, $\vec{x_2} = \eta_3 \vec{q_2}$ and take the integrals over $\vec{q_2}$. As the result we obtain:

\begin{equation}
\begin{split}
n_p(\eta) \approx \frac{8\lambda^2 S_{d-2}}{(2\pi)^{2(d-1)}} \int^{min[\eta^2, \mu^2/p^2]}_{\eta_0^2} \frac{du}{u} \int dx_1 \int \frac{dv}{v^{\frac{d+1}{2}}} \Big[|A_{+}|^2 v^{-i\mu} + |A_{-}|^2 v^{i\mu} \Big] \Big[\bar{h}(x_1) h(x_1 v^{-1}) \Big]^2
\end{split}\label{nLoop}
\end{equation}
and

\begin{equation}
\begin{split}
\kappa_p(\eta) \approx -\frac{16\lambda^2 S_{d-2}}{(2\pi)^{2(d-1)}} \bar{A}_{+} \bar{A}_{-} \int^{min[\eta^2,\mu^2/p^2]}_{\eta_0^2} \frac{du}{u} \int dx_1 \int \frac{dv}{v^{\frac{d+1}{2}}} \Big[v^{-i\mu} + v^{i\mu} \Big] \Big[ \bar{h}(x_1) h(x_1 v^{-1}) \Big]^2,
\end{split}\label{kLoop}
\end{equation}
where $S_{d-2}$ is the volume of the $(d-2)$-dimensional sphere and $min[\eta^2, \mu^2/p^2]$ means the smallest value among $\eta^2$ and $\mu^2/p^2$.

There are two possible limits for the integration over $du$: for $p\eta < \mu$ the contributions to $n_p$ and $\kappa_p$ are proportional to $\ln{( \eta / \eta_{0})}$, while for $p\eta > \mu$ they are proportional to $\ln{(\mu / p \eta_{0})}$. In both situations we cannot take $\eta_0$ to the past infinity due to the explicit infrared divergence in $n_p$ and $\kappa_p$. It is not very hard to trace that one encounters such divergences in CPP for the same reason as why loop corrections in EPP do grow with time (see e.g. \cite{Akhmedov:2013vka} for a review). To cut the divergence in the loop integrals one has to keep the initial Cauchy surface at $\eta_0$, which explicitly breaks the dS isometry. 
In the EPP the initial Cauchy surface can be safely taken to the past infinity ($\eta_0 \to \infty$ in the EPP) because due to such a shift of the surface every physical momentum on it experiences a blue shift. Meanwhile the highly blue shifted BD modes behave as in flat space-time and do not sense the background geometry, as it should be expected on general physical grounds. But in the CPP the movement towards the past infinity results in the infrared shift of the physical momentum. Hence, for the same reason as why one gets secular growth in the EPP one obtains the secular infrared divergence in the CPP. Note that this divergence is present for any value of mass. 

At the same time, the situation in the global dS is similar to the one in the CPP if one restricts attention to the infrared divergences. In fact, the global dS contains simultaneously the CPP and EPP. {An approach to this IR problem typically employed in literature is just to calculate all loop corrections in euclidean dS space with subsequent analytic continuation to lorentzian signature \cite{Marolf:2010zp, Marolf:2010nz,Marolf:2012kh}. One can think of this as of preparing the interacting theory in a maximally analytic state akin to the BD one. However, we cannot use euclidean theory at all for non-equilibrium processes for a generic initial state, so the study of lorentzian situation is also very important. The IR divergences we encountered then mean that the theory is unstable under small time-dependent perturbations happening in the past.}

\subsection{Adiabatic switching vs $i\ep$-prescription}
So far we have performed the Keldysh rotation (\ref{KeldRot}) without thinking about $i\ep$-prescription which was relevant in the previous sections. We did not need it as we imposed a hard cutoff at some finite time and calculated only the most divergent part. This is possible to do in the Schwinger-Keldysh approach. However, as it was pointed out in e.g. \cite{Kaya:2018jdo} and as we discuss in appendix \ref{massMink}, there is a different softer regularization scheme. Instead, one can modify the modes as
\begin{equation}
h\to h_\ep = e^{\ep\tau}h\label{hep}
\end{equation}
and integrate along the entire time line without any past cutoff. Note that it is only necessary to regularize in $\tau \to -\infty$ region as in Schwinger-Keldysh formalism we always have a natural upper cutoff for time integration. Two approaches, with the sharp and soft cutoffs are possible in non-stationary situations with the use of the Schwinger-Keldysh technique. They describe two different physical situations and have different consequences. Wedo our best to consider both situations separately.

{In \cite{Kaya:2018jdo} it was shown that in EPP regularization (\ref{hep}) is equivalent to using the standard $i\ep$-prescription like the one shown on fig. \ref{PropCuts}. However, the situation is different in CPP. The free Wightman function in terms of the modes is expressed as follows:
\begin{equation}
    W_0(p|\eta_1,\eta_2) = (\eta_1\eta_2)^{\frac{d-1}{2}} h(p\eta_1) \bar{h}(p\eta_2).
\end{equation}
For regularization of the time integrals we only care about regions in the past, so we assume $\tau_j < 0$ with $\eta_j = e^{\tau_j}$ as usual. Then the standard prescription $\tau_1-\tau_2-i\ep$ can be achieved if we shift $\tau_1 \to \tau_1 + i\ep \tau_1$, $\tau_2 \to \tau_2 - i\ep \tau_2$. If the modes have only positive energy exponentials, $h \sim e^{-i\alpha\tau}$ with $\alpha > 0$, then such shifts are clearly equivalent to (\ref{hep}). However, from (\ref{HankExp}) we see that this is not the case in CPP as $h \to A_+ e^{i\mu\tau} e^{-\ep \tau} + A_- e^{-i\mu\tau} e^{\ep \tau}$. The first term has ill-defined behavior at past infinity, so the regularization based on analytic properties is not suitable. This is to be expected: in EPP the BD modes diagonalize the hamiltonian $H$ at past infinity and have positive energy \cite{Akhmedov:2013vka}, so we can use the standard Minkowski space trick with shifting $\tau\to \tau + i\ep \tau$ in the expression $e^{-i H \tau}$ in the past to select the interacting vacuum \cite{Peskin:1995ev}. In CPP, however, the free hamiltonian is not diagonalized by the BD modes, so such a shift does not have a clear physical meaning.
}

{Nevertheless, the regularization (\ref{hep}) still admits a physical interpretation. Namely, instead of assigning $e^{\ep\tau}$ factor to the modes we can assign it to the interaction term $\mt{L}_{\text{int}}$ in the lagrangian:
\begin{equation}
    \mt{L}_{\text{int}}\to \mt{L}_{\text{int}} e^{\ep\tau}\label{Lep}
\end{equation}
It simply means that the interaction is slowly turned on starting from the time $\tau \sim -\frac{1}{\ep}$. It is a well-known fact that for usual stationary QFT this adiabatic approach yields the same results as the analytic $i\ep$-prescription \cite{Kamenev,Kamenev2}. As in our case the latter is inapplicable, we will stick to this approach. Besides, it will allow us to practically resum all loops in the next subsection by computing the exact modes.}

{Next, while the analysis of \cite{Kaya:2018jdo} used the <<$+$>> and <<$-$>> representation of the diagrams, the Keldysh rotation described in the above subsection is just its linear transformation. Hence it should yield the same results as long as we use the regularization (\ref{Lep})~--- note that then we always integrate over real $\tau$ without any imaginary shift. Such a prescription makes integrals of oscillating exponentials convergent at $-\infty$ and well-defined in the limit $\ep \to 0$ (we will return to this in the next subsection). However, if there is no oscillating part the limit is still divergent: $\int_{-\infty}^0 e^{\ep \tau} =\frac{1}{\ep}$. This is exactly what we get in (\ref{nLoop}) and (\ref{kLoop}) if we use this prescription instead of the hard cutoff and change $u=e^{\tau}$. Hence the divergent part was computed correctly in the previous subsection. There were also oscillating exponentials of the form $u^{\pm 2i\mu}$ which we did not include in (\ref{nLoop}) and (\ref{kLoop}). The corresponding integrals are technically cutoff dependent but bounded in the case of the hard cutoff, but there are no problems in the adiabatic regularization scheme.}

{Finally, as this regularization is not connected with analytic properties, we cannot expect any resemblance with the euclidean theory. This is precisely what we observed: in euclidean signature there are no IR divergences.}

\subsection{Tree-level infrared divergences in the contracting Poincare Patch} \label{TreeDiv}

The infrared divergences for the BD state discussed in the previous subsection are, in fact, not a property just of loop corrections. In this subsection we will show that somewhat similar problems for the BD state are present even on the tree-level, but they can be solved by choosing a different Hadamard initial state, which however violates the dS isometry on the tree-level. 

To show the origin of the tree-level divergence, let us consider the scalar theory with the following lagrangian:
\begin{equation}
\mt{L} = \dfrac{1}{2}\Big[ g^{\mu\nu}\p_\mu \phi \p_\nu \phi - m^2 \phi^2 - a f(\tau) \phi^2 \Big], \label{mlag}
\end{equation}
where $f(\tau) \ge 0$ is a function which slowly changes from $f(\tau \to -\infty) \to 0$ at past infinity to $f(\tau) \to 1$ at $\tau \approx -\Lambda$, where $\Lambda \gg 1$ has a meaning of the infrared cutoff. It is responsible for the adiabatic switching described in the previous subsection. It is convenient to define this $\Lambda$ as follows:
\begin{equation}
\Lambda = \ili_{-\infty}^0 f(\tau) d\tau.\label{Lambda}
\end{equation}
For instance, we can use $f(\tau) = e^{\gamma \tau}$ if we wish to consider the evolution till the point $\tau_0$ such that $|\tau_0| \gamma \ll 1$ so $e^{\gamma\tau_0} \approx 1$. In this case $\Lambda = \frac{1}{\gamma}$. 

Note that within the Schwinger-Keldysh technique there is no particular physical restriction to change the mass adiabatically. Namely, in Minkowski space we use adiabaticity to keep the system in the stationary Poincare-invariant ground state, while the dS states that we consider are not stationary in usual sense. However, we can think of the maximally-analytic BD state as an analog of Minkowski vacuum in the sense that in it all correlation functions depend only on the geodesic distances rather than on each their argument separately and have well-defined analytic properties. So it is natural to check whether or not the maximal analyticity is preserved under the adiabatic change of the mass term. Besides that, in appendix \ref{massMink} we discuss that the instant change of the mass term causes even more sever peculiarities.

Let us first consider the perturbation theory in $a$. 
After the Keldysh rotation of the fields (\ref{KeldRot}) the lagrangian takes the following form:
\begin{equation}
    \mt{L} = g^{\mu\nu} \p_\mu \phi_{cl} \p_\nu \phi_q - m^2 \phi_{cl} \phi_q - a f(\tau) \phi_{cl} \phi_q.
\end{equation}
Treating the last term as the vertex, we can compute the first-order correction to the Keldysh propagator:
\begin{equation}
    D^K_1(p|\eta_1,\eta_2) = a\ili_0^{+\infty} \dfrac{d\xi}{\xi^D} f(\log \xi) \Big[ D_0^R(p|\eta_1,\xi)D_0^K(p|\xi,\eta_2) + D_0^K(p|\eta_1,\xi)D_0^A(p|\xi,\eta_2) \Big]. \label{DKIR1}
\end{equation}
Here we assume that $|\log \eta_{1,2}| \ll \Lambda$. Due to the presence of $\theta$-functions in $D_0^{R/A}$ the upper limit of integration is actually $\eta_1$ for the first summand and $\eta_2$ for the second one. Note that a finite change of the upper limit shifts the integral by a finite value when $\Lambda \to \infty$. As we are only interested in the most divergent contribution $D^K_{1,\text{IR}}$, we can for simplicity fix it at, say, $\xi = 1$ and forget about $\theta$-functions. Using the expressions for the propagators via the modes (\ref{PropDef}), we find that the leading infrared contribution to $D^K_1$ is contained in the expression:
\begin{equation}
D^K_1(p|\eta_1,\eta_2) \approx a (\eta_1\eta_2)^{\frac{D-1}{2}} \ili_{-\infty}^0 d\tau f(\tau) \left[\dfrac{h(p\eta_1)\bar{h}^2(pe^{\tau})h(p\eta_2)}{i} + \text{c.c.} \right],\label{DKIR}
\end{equation}
where we have made the change of variables as follows: $\xi = e^{\tau}$.

As the divergence may occur at $\tau \to -\infty$, we can use the asymptotic expansion for $h(p\xi)$ from (\ref{HankExp}). Note that in the discussion of this subsection so far we did not explicitly used the fact that $h(p\xi)$ are BD modes: (\ref{HankExp}) can be used for any choice of modes. The only nontrivial condition comes from the canonical commutation relations \cite{Akhmedov:2019esv}:
\begin{equation}
    |A_-|^2 - |A_+|^2 = \dfrac{1}{2\mu}.\label{comrel}
\end{equation}
For instance, in the case of BD modes these coefficients are as follows:
\begin{equation}
    A_-^{\text{BD}} = \dfrac{\sqrt{\pi} e^{\frac{\pi\mu}{2}}}{2^{1+i\mu}\Gamma(1+i\mu)\sinh \pi\mu},\quad A_+^{\text{BD}} = -\dfrac{\sqrt{\pi} e^{-\frac{\pi\mu}{2}}}{2^{1-i\mu}\Gamma(1-i\mu)\sinh \pi\mu}.
\end{equation}

Returning to our situation, not all contributions from the expansion (\ref{HankExp}) produce divergent terms in (\ref{DKIR}). Namely, the integral of $f(\tau) e^{i \alpha \tau}$ is finite and well-defined when $\Lambda \to \infty$ for $\alpha \neq 0$ due to the properties of $f(\tau)$. For example,
\begin{equation}
    \lim_{\gamma\to 0} \ili_{-\infty}^0 d\tau e^{\gamma\tau} e^{i\alpha\tau} = \dfrac{1}{i\alpha}.\label{SoftCut}
\end{equation}
(This property is definitely not the case if $f$ changes rapidly, e.g. if $f(\tau) = 0$ for $\tau < -\Lambda$ and $f(\tau)=1$ for $\tau > -\Lambda$.) Therefore, to pick up the leading divergence we only need to consider the terms without oscillating exponentials. Using (\ref{Lambda}) we find:
\begin{equation}
    D^K_1 \approx 2a\Lambda (\eta_1\eta_2)^{\frac{D-1}{2}} \left[\dfrac{h(p\eta_1)h(p\eta_2) \bar{A}_+ \bar{A}_-}{i} + \text{c.c.} \right].\label{DK1}
\end{equation}
It is evidently divergent when $\Lambda \to \infty$. In the last expression we simply picked out the most singular term of (\ref{DKIR}) and (\ref{DKIR1}).

It is interesting that according to (\ref{1-loop}) this correction corresponds to the generation of a non-trivial anomalous average $\kappa_p = \frac{2\ep\Lambda \bar{A}_+\bar{A}_-}{i}$ with respect to the initial state. Of course the mass of the state  is changed by $a$ when $\tau > -\Lambda$, but we can take $a \to 0$, $\Lambda \to \infty$ in such a way that $a \Lambda = \const$. Effectively it means that in the CPP the BD state acquires a finite anomalous average and, hence, is destroyed by an infinitesimal change of the mass if the cutoff is taken to infinity. 

The only alpha state which has the immunity to this effect is the one with $A_+ = 0$. The modes then are given by $h(x) = \sqrt{\frac{\pi}{\sinh \pi \mu}} J_{-i\mu}(x)$ which corresponds to the In-~state in CPP \cite{Akhmedov:2019esv}. However, it is an alpha state with $\alpha \neq 0$ and therefore has UV problems in perturbation theory, which were discussed in the subsection \ref{EPPSP}. 

It is worth noting that our consideration is not always applicable if the cutoff is finite. Namely, it can be used if the expansion of $h(p\xi)$ from (\ref{HankExp}) is valid near the cutoff scale, i.e. when $\xi \sim e^{-\Lambda}$. It means that $p \ll \mu e^{\Lambda}$. On the other hand, when $p \gg \mu e^{\Lambda}$, we need to use a different asymptotic expansion:
\begin{equation}
    h(x) = B_+ \dfrac{e^{ix}}{\sqrt{x}} + B_- \dfrac{e^{-ix}}{\sqrt{x}},~x \gg \mu.
\end{equation}
For the BD state $B_+ = 0$ and, hence, $\bar{h}^2(p\xi)$ contains only oscillating terms. 

In all, the modes with high enough momentum do not feel the correction (\ref{DK1}) if the cutoff is finite. It means that if one chooses such modes that behave as In-modes in the CPP when $p \ll \mu e^{\Lambda}$ and as the BD modes when $p \gg \mu e^{\Lambda}$, then correlation functions have the same UV-behavior as BD state and at the same time are not sensitive to a small change of the mass term that we are discussing in this subsection. It is possible to choose such modes by considering a linear combination of the BD modes with momentum-dependent coefficients. Namely, one can choose as the mode functions the following canonically transformed expressions:
\begin{equation}
g_p(\eta, \vec{x}) = \eta^{\frac{D-1}{2}} \, \int d^{D-1}q \Big[\alpha_{pq} \, h(q\eta) \, e^{i\vec{q}\vec{x}} + \beta_{pq} \, \bar{h}(q\eta) \, e^{-i\vec{q}\vec{x}}\Big],
\end{equation}
where the complex $\eta$-independent coefficients $\alpha$ and $\beta$ should obey certain conditions to fulfil canonical commutation relations for the field operator and the corresponding conjugate momentum and for the ladder operators. Also one can impose such conditions on $\alpha$ and $\beta$ that the corresponding correlation functions for the Fock space ground state will obey the Hadamard conditions. E.g. one can take $\alpha_{pq} = \alpha_p \delta(\vec{p}-\vec{q})$ and $\beta_{pq} = \beta_p \delta(\vec{p}-\vec{q})$ so the modes $g_p$ are still given in a Fourier basis and interpolate between ones with $A_+ = 0$ for small $p$ and BD modes for large $p$. However, the corresponding correlation functions, which are build with the use of the Fock space ground state for these modes, will explicitly violate the dS isometry already at the tree-level.

{A sumilar calculation for the global dS space was performed in \cite{Marolf:2011sh} in one- and two-loop orders. However, the two propagators in each product in the integrand of the first-order similar to (\ref{DKIR1}) had different mass parameters $\mu_1$ and $\mu_2$. Hence, the terms with oscillating exponentials of either $\mu_1 + \mu_2$ or $\mu_1 -\mu_2$ were always present, so IR divergences did not appear. Besides that, we will now show a rather simple way to resum all of the loops.}

For the lagrangian (\ref{mlag}) the straightforward resummation of the infrared divergences is possible. Let us perform it to conclude this subsection. Instead of dealing with diagrammatic series, it is much easier to explicitly solve the quadratic theory defined by (\ref{mlag}) in the WKB approximation. The equation for exact modes $\hat{h}_{a}(\tau)$ is as follows:
\begin{equation}
    \Big[\p_\tau^2 + p^2 e^{2\tau} + \mu^2 + a f(\tau)\Big] \hat{h}_a(\tau) = 0.\label{modeq}
\end{equation}
Due to the properties of $f(\tau)$ solutions of this equation are modes for the theory with the mass parameter $\mu$ when $\tau \to -\infty$ and with the mass parameter $\sqrt{\mu^2 + a}$ when $|\tau| \ll \Lambda$ or $\tau > 0$. Let us for simplicity perform the computation in the following limit:
\begin{equation}
a\to 0,\quad \Lambda \to \infty,\quad a \Lambda = \const. \label{LimEp}    
\end{equation}
Then the modes when $\tau < 0$, $|\tau|\ll\Lambda$ or when $\tau > 0$ correspond to the theory with the mass parameter $\mu$ as well:
\begin{equation}
    \hat{h}_a(\tau) = \begin{cases}
    h(pe^{\tau}),&\tau \to -\infty\\
    \tilde{h}(p e^{\tau}),&|\tau|\ll\Lambda~\text{or}~\tau>0.
    \end{cases}\label{ModsAs}
\end{equation}
We assume that the modes $h(x)$ are given (i.e. we know $A_+$ and $A_-$ in the expansion of $h$) and we need to find $\tilde{h}$. 

It is more convenient to work in the region where $p e^{\tau} \ll 1$, so (\ref{HankExp}) can be used. We can then neglect the term $p^2 e^{2\tau}$ in (\ref{modeq}). As $f$ changes slowly, the solution of the resulting equation can be found in the WKB approximation:
\begin{equation}
    \hat{h}_{a}(\tau) = C_1 \exp\lb i\ili_{0}^{\tau}\sqrt{\mu^2 + a f(\tau')}d\tau'\rb + C_2 \exp\lb -i\ili_{0}^{\tau}\sqrt{\mu^2 + a f(\tau')}d\tau'\rb.
\end{equation}
Here we preserved $a$-terms only in the exponents, where $a\Lambda$ combination may arise, neglecting them in the preexponential factor $\frac{1}{(\mu^2 + a f(\tau))^{1/4}}$ which also appears in WKB solutions. Let us also expand the square roots in the exponents to the first order in $a$. When $\tau < 0$ and $|\tau| \gg \Lambda$ we find:
\begin{equation}
    \hat{h}_{a}(\tau) = C_1 e^{-i \delta} e^{i\mu \tau} + C_2 e^{i\delta} e^{-i\mu\tau},\quad \delta = \dfrac{a \Lambda}{2\mu}.
\end{equation}
Here we used that $f(\tau) = 0$ for $\tau < 0$ and $|\tau| \ll \Lambda$, so $\int_{0}^\tau f(\tau') d\tau' = \int_{0}^{-\infty} f(\tau') d\tau' = -\Lambda$. The second-order terms in the square root expansion lead to the correction to the phase $\delta$ of the order $a^2 \Lambda$, which is zero in our limit. Using (\ref{HankExp}) and (\ref{ModsAs}) we find:
\begin{equation}
C_1 = A_+ p^{i\mu} e^{i\delta},\quad C_2 = A_- p^{-i\mu} e^{-i\delta}.\label{C1C2}
\end{equation}

On the other hand, when $\tau < 0$, $|\tau| \ll \Lambda$ (it intersects with the region $p e^{\tau} \ll 1$ for sufficiently large $\Lambda$) we have:
\begin{equation}
    \hat{h}_{a}(\tau) = C_1 e^{i\mu \tau} + C_2 e^{-i \mu \tau},
\end{equation}
as $a \int_0^\tau f(\tau') d\tau'$ can be neglected after (\ref{LimEp}). Assuming that $\tilde{A}_+$ and $\tilde{A}_-$ are coefficients in the expansion (\ref{HankExp}) of $\tilde{h}$ and using (\ref{C1C2}) we obtain:
\begin{equation}
\tilde{A}_+ = A_+ e^{i\delta},\quad \tilde{A}_- = A_- e^{-i\delta}.
\end{equation}
In particular, it means that
\begin{equation}
    \tilde{h}(x) \approx \dfrac{|A_+|^2 e^{i\delta} - |A_-|^2 e^{-i\delta}}{|A_+|^2 - |A_-|^2}h(x) + \dfrac{2i A_+ A_- \sin \delta}{|A_-|^2 - |A_+|^2}\bar{h}(x),
\end{equation}
as this holds for asymptotic expansions, and the relation of $\tilde{h}$ with $\bar{h}$ and $h$ should be linear. Using (\ref{comrel}) we can simplify the last expression to:
\begin{equation}
    \tilde{h}(x) \approx \lb e^{-i\delta} - 4i\mu |A_+|^2 \sin \delta \rb h(x) + 4i\mu A_+ A_- \sin \delta\,\bar{h}(x).\label{tilh}
\end{equation}
We can use this relation to express the propagators $\tilde{D}^{K/R/A}$ (\ref{PropDef}) for the modes $\tilde{h}$ in terms of $D_0^{K/R/A}$ for the modes $h$. Retarded and advanced propagators are defined in terms of the field commutator, so they do not depend on the choice of the basis of modes and stay the same, while the Keldysh propagator acquires the form:
\begin{equation}
\begin{aligned}
\tilde{D}^K(p|\eta_1,\eta_2) &\approx D^K_0(p|\eta_1,\eta_2)+\\
&+(\eta_1 \eta_2)^{\frac{D-1}{2}}\left[8\mu|A_+|^2(1+2\mu|A_+|^2)\sin^2\delta \,h(p\eta_1) \bar{h}(p\eta_2)  -\right.\\
&-\left. 4\mu\bar{A}_+ \bar{A}_-\sin\delta \lb i  e^{-i\delta} + 4\mu|A_+|^2\sin\delta\rb h(p\eta_1)h(p\eta_2)+ \text{c.c.} \right].
\end{aligned}
\end{equation}
We see that infrared divergence turned into the phase $\delta$, which is a rather standard phenomenon for such situations. When $\delta \to 0$ one can easily see that the perturbative answer (\ref{DK1}) is reproduced.

By finding the exact modes we effectively resummed all orders of perturbation theory. Hence this computation shows that IR divergences do not completely disappear even after such a resummation if $A_+ \neq 0$. While the result is no longer divergent, there is no well-defined limit $\Lambda \to \infty$ as the cutoff scale only enters the phase terms. On the other hand, the limit that makes sense is $\Lambda \to \infty$, $a \to 0$, $a \Lambda = \const$ and the corresponding correction to the Keldysh propagator does not vanish in this limit. It means that the initial state acquires a non-vanishing particle density and anomalous averages under an infinitesimal shift of the mass term as long as $A_+ \neq 0$.

Hence, we once again find that the BD state in the CPP is unstable and we need to explicitly break the dS isometry symmetry by introducing some form of cutoff and changing low-momentum modes to improve the situation. {Besides that, note that this consideration is somewhat universal. Namely, if we have a theory with the modes behaving as (\ref{HankExp}) at past infinity, i.e. satisfying the wave equation with mass parameter $\mu$, we can reconstruct the WKB solutions to it in the same way and obtain (\ref{tilh}). This is the case in e.g. global dS space, so just as we noted in the subsection \ref{1-loopP} the global dS space also suffers from the IR divergences.}

\section{Conclusions}

We consider loop corrections to tree-level correlation functions in various patches (regions, wedges or charts) of Minkowski and de Sitter space-times for the isometry invariant states. Namely in Minkowski space-time we consider Poincare invariant state, while in de Sitter space-time we mainly consider the Bunch-Davies state. We essentially use the consideration of various patches of Minkowski space-time as model examples for the study of the patches in de Sitter space-time.

Our main goal is to find out if the loop corrections respect the isometries or not and to check if the loop integrals can be mapped to the euclidean signature via analytical continuation of the propagators and rotations of the integration contours in the loop corrections. By product we prove that the Feynman technique does not provide correct answers for the loop integrals, if one integrates in the vertices over the regions rather than over the entire space-times. That is true even if one restricts the consideration to the static charts. {Our consideration was rather general; in the case of de Sitter space it was pointed out in \cite{Higuchi:2008tn,Higuchi:2009ew}. }Furthermore, we prove that the Schwinger-Keldysh technique is causal: the integration over the vertices in the loops can be restricted to the union of the past light cones of the external vertices of the correlation functions. The integrals beyond the past light-cones provide vanishing contributions. {This fact is mentioned in e.g. \cite{Higuchi:2010xt,Polyakov:2012uc} without a formal proof, so we find it instructive to provide one.} We use these observations to prove our statements.

Using method similar to one of \cite{Higuchi:2010xt}, we prove that for the Poincare invariant state the loop corrections in the left and right Rindler wedges of Minkowski space-time can be mapped to the integrals over the euclidean space. Hence, loop corrections in such cases respect the Poincare isometry. {Besides that, we give a simple proof of isometry-invariance which is not related to analytic continuation. While a similar result was proven in \cite{Higuchi:2020swc}, it used an explicit momentum-space representation. We use a more general coordinate-space approach which is based solely on analytic properties and provides a simpler proof.} The same is true for the past or lower wedge. However, in the future wedge loop corrections do not respect the isometry. 

These all observations are heavily relying on the analytic properties of the propagators and on the causality of the Schwinger-Keldysh technique. Thus, obviously although we consider the same tree-level invariant propagators loop corrections still depend on the choice of the patch. Namely,
loop corrections strongly depend on the choice of the geometry of the initial Cauchy surface. In other words they depend on the choice of the initial state. That is due to infrared effects, which are sensitive to the initial and/or boundary conditions. 

Similar situation we encounter in the patches of de Sitter space-time. Namely, for the Bunch-Davies state (for the massive scalar field) we show that loop corrections in the static patch and in the the expanding Poincare patch respect the isometry and can be mapped to the calculation on the sphere. {While the analysis of the static patch case is close to the one of \cite{Higuchi:2010xt}, we prove the EPP isometry-invarince independently using arguments similar to those in \cite{Polyakov:2012uc, Akhmedov:2013vka}.} Then we show that for generic alpha states loop corrections violate the de Sitter isometry. That is related to the analytic properties of the propagators for generic alpha states as opposed to those for the Bunch-Davies state.

Furthermore, we show that loop corrections for the Bunch-Davies state in the contracting Poincare patch and in global de Sitter space-time contain infrared divergences. Namely, even after the subtractions of the UV divergences loop integrals are infinite, if the initial Cauchy surface is placed at past infinity. {They are also present for all values of the mass, unlike the usual IR divergences studied in \cite{Serreau:2013psa, LopezNacir:2016gzi, Gautier:2015pca, Beneke:2012kn,Hollands:2011we}. Our results include 1-loop corrections studied before in \cite{Polyakov:2012uc,Marolf:2011sh}, but we also compute an exact correction for $\phi^2$-theory.} To remove the divergence one has to keep the initial Cauchy surface at a finite initial time $t_0$. But such a cutoff violates the de Sitter isometry, because there are generators of the latter symmetry that can move the position,  $t_0$, of the initial Cauchy surface. Another way considered in the literature is based on the analytical continuation of the euclidean theory \cite{Marolf:2010zp, Marolf:2010nz,Marolf:2012kh}, but it is not suitable for a generic  non-equilibrium initial states. And, as we see, does not lead to the same result obtained in the lorentzian signature for adiabatic evolution in the future Minkowski wedge, contracting Poincare patch and global de Sitter space-time.

Moreover, we show that in the contracting Poincare patch and in global de Sitter space-time one has to consider initial non-invariant Hadamard states to avoid some (but not all) of the infrared problems at every perturbative order. To study the destiny of the other infrared contributions one has to resumm at least the leading loop corrections. The physical reasons of the violation of the isometry in the loops in the patches of de Sitter space-time are discussed in \cite{Akhmedov:2019cfd} in detail. Briefly speaking, in the simplest situations the violation of the isometry appears due to the fact that in non-stationary situations level populations and anomalous averages are changing in time, while in stationary situations they remain zero. 

The violation of the dS isometry at the loop level potentially may have strong physical consequences. In fact, if the isometry is not broken then the expectation value of the stress-energy tensor of the quantum feild theory on the background in question is proportional to the metric tensor with the constant coefficient of proportionality. Then the result of the backreaction of quantum fluctuations is just a renormalization of the cosmological constant. However, if the isometry is broken then in the renormalization one can potentially obtain a screening in time of the cosmological constant, which is similar to the screening of strong electric fields in QED. The latter may happen due to the generation of non-trivial stress-energy fluxes caused the change in time of level populations and anomalous averages.

\section{Acknowledgements} We would like to thank K.Bazarov, F.Popov, A.Polyakov, U.Moschella and especially to D.Diakonov for valuable discussions. The work of ETA and MNM was funded by the Russian Science Foundation (Grant No.23-22-00145).

\begin{appendices}
\section{The corrections to the propagator in the future wedge and in the Minkowski space}\label{propcalc}
We wish to compute the integrals (\ref{FM}) and (\ref{FF}). Let us start from the first one. As the integration goes over the whole Minkowski space-time, we can use the analytic continuation to the euclidean space $\mathbb{R}^4$ with $Y^0 = -iY^4_E$ to obtain:
\begin{equation}
\begin{aligned}
F_E^{(1)}(0,X_E) &= -m^2\ili d^4 Y_E \dfrac{1}{Y_E^2(Y_E-X_E)^2} = -m^2 \ili_0^1 du \ili d^4 Y_E \dfrac{1}{\left[(1-u)Y_E^2 + u (Y_E-X_E)^2 \right]^2} = \\
& = -\pi^2 m^2 \ili_0^1 du \ili_0^{+\infty} \dfrac{v dv}{[v+u(1-u)X_E^2]^2} = -\pi^2 m^2 \lb \log \dfrac{\Lambda^2}{X_E^2} - 1\rb = -\pi^2 m^2 \log \dfrac{\Lambda^2}{X_E^2},
\end{aligned}
\end{equation}
here $\Lambda$ is an IR cutoff scale. Here we have made the following changes: $Y_E\to Y_E - u X_E,~Y_E^2 = v$ and $\Lambda \to e^{-1/2}\Lambda$. As we use in Minkowski space-time such a signature that space-like intervals have negative squares, we find that:
\begin{equation}
F_M^{(1)}(0,X) = -\pi^2 m^2 \log \dfrac{\Lambda^2}{-X^2} = -\pi^2 m^2 \log\dfrac{\Lambda^2}{-t^2}.
\end{equation}

Now let us move on with the calculation in the future wedge of the integral (\ref{FF}). It is convenient to represent the $\delta$-function as follows:
\begin{equation}
\delta[(Y-X)^2] = \dfrac{1}{2|Y^0-t|}\Big[\delta(Y^0-r-t) + \delta(Y^0+r-t)\Big],
\end{equation}
where $r^2 = (Y^i)^2$, $i=1,\dots,3$. Note that the first $\delta$-function on the RHS is always zero as in the integration region $Y^0 < t$. It is convenient to switch from $Y^i$ to spherical coordinates with angles $\theta,~\phi$ and $Y^1 = r \cos\theta$. Then we have:
\begin{equation}
\begin{aligned}
F^{(1)}_F(0,X) &= 2\pi m^2 \ili_{|Y^1|<Y^0<t} d^4Y \dfrac{\delta[(Y-X)^2]}{(Y^2-i\ep)} = 2\pi^2 m^2 \ili_{r|\cos\theta| < t-r} dr\, d(\cos\theta)\dfrac{r^2 }{r(t^2-2tr-i\ep)} = \\
&=2\pi^2 m^2 \left[ 2\ili_0^{t/2} dr \dfrac{r}{t^2-2tr-i\ep} + \ili_{t/2}^t dr \dfrac{r}{t^2-2tr-i\ep} \ili_{-(t-r)/r}^{(t-r)/r} d(\cos\theta) \right].
\end{aligned}
\end{equation}
In the second integral one can change $r\to t-r$ to obtain that
\begin{equation}
 F^{(1)}_F(0,X) = 8i\pi^2 m^2 \ili_0^{t/2}dr\dfrac{r \ep}{\ep^2 + t^2(t-2r)^2}=i\pi^3 m^2.
\end{equation}

\section{Pauli-Villars regularization}\label{PVreg}

In this paper we use the Pauli-Villars regularization to deal with UV divergences in the loop integrals. I.e. we introduce several massive fields to replace the original Green function by the regularized one of the form:
\begin{align}\label{reg}
    \widetilde{G} (\zeta) = G(\zeta, m) + \sum^{[D/2]}_{i=1} \alpha_i G(\zeta, m_i),
\end{align}
where $\alpha_i$ are constants, $m_i$ are masses of additional fields and $\zeta$ is the dS or Poincare invariant, depending on whether we are dealing with dS or Minkowski space-time.

For the BD state in dS space-time the propagator $G(\zeta)$ has a singularity at $\zeta=1$, which is the standard UV divergence.  In the limit $\zeta \to 1$ the BD propagator can be represented as:
\begin{align}
\label{Gexp}
G(\zeta, m_i)=\sum_{n=0}^{[D/2]-1} \frac{g_{n}(m_i)}{(1-\zeta)^{\frac{D-2}{2}- n}}+g(m_i) \log(1-\zeta) + f(\zeta),
\end{align}
where $g_{n}(m_i)$ and $g(m_i)$ are constants that depend on the masses and Hubble constant, $f(\zeta)$ is a de Sitter invariant bounded function that has well-defined finite limit as $\zeta\to 1$ and branch cut from $1$ to $ +\infty$. The logarithmic term in (\ref{Gexp}) is present only in even dimensions.  

Hence to make $\widetilde{G} (\zeta) $ well-defined finite function in the limit $\zeta\to 1$, $\alpha_i$ should obey the following relations:
\begin{align}
g_{n}(m)+\sum_{i=1}^{[D/2]}\alpha_i g_n(m_i)=0 ,\quad  \text{for} \quad n=0, 1, ...,[D/2]-1\quad 
\end{align}
and 
\begin{align}
g(m)+\sum_{i=1}^{[D/2]}\alpha_i g(m_i)=0 .
\end{align}

After this regularization  the propagator $\widetilde{G}$ is still dS invariant, since each propagator in \eqref{reg} is separately dS invariant, and has the same analytic properties in the complex $\zeta$-plane as the original propagator. We suppose that $m_i \gg 1$ in the units of the dS curvature. In this paper we assume that all propagators are regularized in the described here way.

\section{Change of the mass term in Minkowski space-time}\label{massMink}
To illustrate the situation with switching on the mass term in CPP on a simple example in this appendix we consider the case when the mass is changed rapidly in Minkowski space-time and compare it with the adiabatic case. The modes in Minkowski space-time are just the standard plane waves, but we choose them mixed:
\begin{equation}
g_p(t) = B_+ \dfrac{e^{i\w_p t}}{\sqrt{2\w_p}} + B_- \dfrac{e^{-i\w_p t}}{\sqrt{2\w_p}},\quad \w_p = \sqrt{m^2 +p^2},\quad |B_-|^2 - |B_+|^2 = 1.\label{MinkModes}
\end{equation}
The last condition is needed to satisfy the canonical commutation relations. Of course the standard choice of positive energy modes is $B_+ = 0$, $B_- = 1$, but the situation with non-zero $B_+$ mimics the one we encounter for the BD modes in the CPP. 

The Wightman function for the Fock space ground state for these modes is as follows:
\begin{equation}
    W_g(X,X') = |B_-|^2 W(X,X') + |B_+|^2 W(X_t,X_t') + B_- \bar{B}_+ W(X,X'_t)+ \bar{B}_- B_+ W(X_t,X').
\end{equation}
Here $X_t$ is $X$ with the reversed time: $X_t^0 = -X^0$, $X_t^i = X^i$ and $W(X,X')$ is the standard Wightman function for the scalar field theory with mass $m$. From here it is clear that such states are not Poincare-invariant\footnote{Furthermore, these propagators do not obey the proper UV Hadamard behaviour. To obtain the propagator with the proper Hadamard behaviour one should take such $B_\pm$ which depend on $p$ and $B_+(p) \to 0$ as $|p| \to \infty$.} if $B_+ \neq 0$. We will still consider them, however, to compare with the CPP case.

Now let us assume that at $t = -\Lambda$ the mass is instantaneously changed as follows: $m^2\to \tilde{m}^2$. It means that the exact modes $\hat{g}_p$ satisfy the following equation:
\begin{equation}
    \Big[\p_t^2 + p^2 + m^2 + (\tilde{m}^2-m^2) \theta(t+\Lambda)\Big]\hat{g}_p(t) = 0.
\end{equation}
Assuming that $g(t)$ is defined by (\ref{MinkModes}) when $t < -\Lambda$, one can easily find the modes when $t>-\Lambda$:
\begin{equation}
\begin{aligned}
\hat{g}_p(t) &=  \tilde{B}_+ \dfrac{e^{i\tilde{\w}_p t}}{\sqrt{2\tilde{\w}_p}} + \tilde{B}_- \dfrac{e^{-i\tilde{\w}_p t}}{\sqrt{2\tilde{\w}_p}};\\
\tilde{B}_+ &= \dfrac{B_+ (\w_p +\tilde{\w}_p) e^{i\Lambda (\tilde{\w}_p - \w_p)} + B_-(\tilde{\w}_p - \w_p)e^{i\Lambda(\tilde{\w}_p+\w_p)}}{2\sqrt{\w_p\tilde{\w}_p}};\\
\tilde{B}_-&= \dfrac{B_- (\w_p +\tilde{\w}_p) e^{-i\Lambda (\tilde{\w}_p - \w_p)} + B_+(\tilde{\w}_p - \w_p)e^{-i\Lambda(\tilde{\w}_p+\w_p)}}{2\sqrt{\w_p\tilde{\w}_p}},
\end{aligned}
\end{equation}
here $\tilde{\w}_p = \sqrt{\tilde{m}^2+p^2}$. Note that the first summands in the equations defining $\tilde{B}_+$ and $\tilde{B}_-$ are expressions which appear in the WKB approximation (although with different pre-exponential factors), while the second ones do not appear in the WKB approximation. In particular, even in the case of invariant vacuum $B_+ = 0$ we have $\tilde{B}_+ \neq 0$. It is not surprising, because an instant change of the mass term should move the theory away from the vacuum state.

Now let us find the Keldysh propagator. It is convenient to express $\hat{g}_p(t)$ when $t > -\Lambda$ in terms of the modes $\tilde{g}_p$ defined for the mass $\tilde{m}$ with $B_+$ and $B_-$ being the same:
\begin{equation}
    \tilde{g}_p(t) ={B}_+ \dfrac{e^{i\tilde{\w}_p t}}{\sqrt{2\tilde{\w}_p}} + {B}_- \dfrac{e^{-i\tilde{\w}_p t}}{\sqrt{2\tilde{\w}_p}}.\label{tmModes}
\end{equation}
We find:
\begin{equation}
\begin{aligned}
    \hat{g}_p(t) &= \dfrac{\tilde{g}_p(t)}{2\sqrt{\w_p\tilde{\w}_p}}\left[ ( \w_p + \tilde{\w}_p )\lb |B_-|^2e^{-i\Lambda(\tilde{\w}_p - \w_p)} - |B_+|^2 e^{i\Lambda(\tilde{\w}_p - \w_p)} \rb+\right.\\
    &+\left.(\tilde{\w}_p - \w_p) \lb B_+\bar{B}_- e^{-i\Lambda (\w_p + \tilde{\w}_p)} - B_-\bar{B}_+ e^{i\Lambda (\w_p + \tilde{\w}_p)}  \rb\right]+\\
    &+\dfrac{\bar{\tilde{g}}_p(t)}{2\sqrt{\w_p\tilde{\w}_p}}\left[ ( \tilde{\w}_p-\w_p )\lb B_-^2e^{i\Lambda(\tilde{\w}_p + \w_p)} - B_+^2 e^{-i\Lambda(\tilde{\w}_p + \w_p)} \rb+\right.\\
    &+\left.(\w_p+\tilde{\w}_p)B_+B_- \lb e^{i\Lambda(\tilde{\w}_p - \w_p)} - e^{-i\Lambda(\tilde{\w}_p - \w_p)}\rb \right].
\end{aligned}
\end{equation}
Now the Keldysh propagator at $t > -\Lambda$ can be expressed as follows:
\begin{equation}
    \hat{D}^K(p|t_1,t_2) = \tilde{D}_0^K(p|t_1,t_2) + \Big[\tilde{g}_p(t_1)\bar{\tilde{g}}_p(t_2) n_p + \tilde{g}_p(t_1) \tilde{g}_p(t_2)\kappa_p + \text{c.c}\Big],
\end{equation}
where $\tilde{D}_0^K(p|t_1,t_2)$ is the Keldysh propagator for the modes $\tilde{g}_p$ and
\begin{equation}
\begin{aligned}
n_p &= \dfrac{1}{4\w_p\tilde{\w}_p}\left\{(\tilde{\w_p}-\w_p)^2\left[ 1- (\bar{B}_+^2 B_-^2 e^{2i\Lambda (\w_p+\tilde{\w}_p)} + \text{c.c.}) \right] +\right.\\
&+\left. 2|B_+B_-|^2\left[2(\w_p^2+\tilde{\w}_p^2) - (\w_p+\tilde{\w}_p)^2\cos[2\Lambda(\tilde{\w}_p-\w_p)]\right]-\right.\\
&- \left. (\tilde{m}^2-m^2) \left[(|B_-|^2 + |B_+|^2)\bar{B}_+ B_- (e^{2i\Lambda \tilde{\w}_p} - e^{2i\Lambda \w_p}) + \text{c.c.} \right]\right\};\\
\kappa_p &= \dfrac{1}{4\w_p\tilde{\w}_p} \left\{ (\w_p+\tilde{\w}_p)^2\bar{B}_+\bar{B}_-\lb |B_-|^2e^{-2i\Lambda(\tilde{\w}_p-\w_p)}-|B_+|^2\rb \lb 1- e^{2i\Lambda(\tilde{\w}_p-\w_p)}\rb +\right.\\
&+\left. (\tilde{\w}_p-\w_p)^2\lb B_+\bar{B}_- - B_-\bar{B}^+ e^{2i\Lambda(\w_p+\tilde{\w}_p)}\rb \lb \bar{B}_-^2 e^{-2i\Lambda(\w_p+\tilde{\w}_p) } - \bar{B}_+^2 \rb + \right.\\
&+\left. (\tilde{m}^2-m^2)\left[ (|B_-|^2+|B_+|^2)\lb \bar{B}_-^2 e^{-2i\Lambda \tilde{\w}_p} + \bar{B}_+^2 e^{2i\Lambda \tilde{\w}_p} \rb - 2 \bar{B}_-\bar{B}_+\lb \bar{B}_+B_-e^{2i\Lambda \w_p} + \text{c.c.} \rb \right] \right\}.
\end{aligned}\label{DensInst}
\end{equation}
These expressions define particle density and anomalous averages with respect to the state with the mass $\tilde{m}$ and modes (\ref{tmModes}). Note that they have terms which do not vanish when $\tilde{m}\to m$ and $\Lambda (\tilde{m}^2-m^2) = \const$ and the limit $\Lambda \to \infty$ is undefined.

For comparison let us also find these coefficients in the adiabatic case. In the same way as in the subsection \ref{TreeDiv} we find the relevant modes:
\begin{equation}
\hat{g}_p^{\text{ad}}(t) =  B_+ e^{i\Lambda (\tilde{\w}_p - \w_p)} \dfrac{e^{i\tilde{\w}_p t}}{\sqrt{2\tilde{\w}_p}} + B_- e^{-i\Lambda (\tilde{\w}_p - \w_p)} \dfrac{e^{-i\tilde{\w}_p t}}{\sqrt{2\tilde{\w}_p}};
\end{equation}
Then the particle density and anomalous averages are as follows:
\begin{equation}
\begin{aligned}
n_p^{\text{ad}} &= 4|B_+|^2|B_-|^2\sin^2[\Lambda(\tilde{\w}_p - \w_p)]\\
\kappa_p^{\text{ad}} &= -2i \bar{B}_+\bar{B}_- \lb |B_-|^2 e^{-i\Lambda (\tilde{\w}_p - \w_p)} - |B_+|^2 e^{i\Lambda (\tilde{\w}_p - \w_p)} \rb\sin[\Lambda (\tilde{\w}_p - \w_p)].
\end{aligned}\label{DensAd}
\end{equation}
We see that these expressions are much simpler than those from (\ref{DensInst}). First, all the phases here are proportional to $\Lambda(\tilde{\w}_p-\w_p)$, while in (\ref{DensInst}) some of them are proportional to $\Lambda(\tilde{\w}_p+\w_p)$. From the perturbation theory perspective, it happens because when we integrate $e^{i\alpha x}$ with a hard cutoff at $-\Lambda$ instead of a soft one as in (\ref{SoftCut}) there is a nonzero contribution from the lower limit. Second, $n_p^{\text{ad}}=\kappa_p^{\text{ad}} = 0$ for the invariant vacuum $B_+ = 0$ and $B_- = 1$, which is not the case for $n_p$ and $\kappa_p$. Let us denote the latter by $n_p^-$ and $\kappa_p^-$. They have the following form:
\begin{equation}
n_p^- = \dfrac{(\tilde{\w}_p-\w_p)^2}{4\w_p\tilde{\w}_p},\quad \kappa_p^{-} = \dfrac{\tilde{m}^2-m^2}{4\w_p\tilde{\w}_p}e^{-2i\Lambda \tilde{\w}_p}.
\end{equation}
It is something to be expected as an instant change of the mass term should distort the ground state.

Finally, it is instructive to compare the result with the case of CPP. The modes in the past are still oscillating exponential, so the similar analysis is applicable. However the crucial difference is that the momentum in the equation (\ref{modeq}) is exponentially suppressed, so the frequencies of the modes at the past infinity are momentum-independent ($\mu$ when $\tau < -\Lambda$ and $\tilde{\mu}$ when $\tau > -\Lambda$) as long as $p e^{-\Lambda} \ll \mu$. Hence the phases in CPP computation are proportional to $\Lambda(\tilde{\mu}-\mu)$ and $\Lambda (\tilde{\mu}+\mu)$ (if the mass is rapidly changed). When $\Lambda \to \infty$, the frequencies of all modes become constant.

On the other hand, the momentum dependence of $\w_p$ in Minkowski space becomes relevant when we do the Fourier transformation to the coordinate space. Then we can treat $\hat{D}^K(p|t_1,t_2)$ not as a function of momentum $p$, but as a distribution, so the limit $\Lambda \to \infty$ becomes well-defined in some cases. Namely, the terms with phases proportional to $\Lambda(\w_p+\tilde{w}_p)$, $\Lambda \w_p$ and $\Lambda \tilde{\w}_p$ are suppressed as $\Lambda \to \infty$, as they effectively restrict the momentum integration region to $p \lesssim \sqrt{\frac{m}{\Lambda}}$. For instance, $\kappa_p^-$ vanishes as a distribution, but $n_p^-$ is $\Lambda$-independent. Thus, the invariant vacuum state is still distorted after the rapid change of the mass term even if we take the limit $\Lambda \to \infty$ in the generalized sense. The correction, however, is of the order $(\tilde{m}-m)^2$.

The situation with phases proportional to $\Lambda(\tilde{\w}_p-\w_p)$ is more interesting. Let us consider the following integral in the case $d=4$:
\begin{equation}
    \mt{I}(\vec{x}) = \ili d^3 p\, \dfrac{e^{i\Lambda (\tilde{\w}_p-\w_p)+i\vec{p}\vec{x}}}{\tilde{\w}_p}.
\end{equation}
We will find its asymptotic form when $\Lambda \to \infty$ using the steepest descent method. As we will show, the stationary point is at $|\vec{p}| \gg m$ if $\Lambda$ is sufficiently large (we assume that $\tilde{m}\sim m$), so we can use the relevant expansions of $\w_p$ and $\tilde{\w}_p$. Assuming that $\vec{x} = (x,0,0)$ with $x>0$ and integrating over the angular variables we find:
\begin{equation}
    \mt{I}(\vec{x}) = \dfrac{2\pi}{i x} \lb\ili_0^{+\infty}dp\, e^{\frac{i\Lambda \hat{a}}{p} + ipx} - \ili_0^{+\infty}dp\, e^{\frac{i\Lambda \hat{a}}{p} - ipx}\rb,\quad \hat{a} = \dfrac{\tilde{m}^2-m^2}{2},
\end{equation}
where we assume that $\hat{a} > 0$. It is now straightforward to find stationary points: they are $\pm \sqrt{\frac{\Lambda\hat{a}}{x}}$ for the first and second integral, correspondingly. The second point, however, does not belong to the interval $(0,+\infty)$. Hence, the second integral is suppressed~--- we only need to consider the contribution of $p_0 = \sqrt{\frac{\Lambda\hat{a}}{x}}$. Just as we assumed, $p_0 \gg m$ if $\Lambda$ is sufficiently large, e.g. $x\ll \frac{\hat{a}\Lambda}{m^2}$. We have:
\begin{equation}
    \mt{I}(\vec{x})\approx \dfrac{2\pi} {ix}e^{2i\sqrt{\Lambda\hat{a}x}} \ili_{-\infty}^{\infty} dq\,e^{iq^2\frac{x^{3/2}}{(\Lambda \hat{a})^{1/2}} } = \dfrac{2\pi^{3/2} (\hat{a}\Lambda)^{1/4}}{i^{1/4} x^{7/4}} e^{2i\sqrt{\Lambda \hat{a}x}}.
\end{equation}
Note that this expression is divergent when $\Lambda\to \infty$~--- there is a large region near $p_0$ where the phase is almost constant. Hence, the corrections to the Keldysh propagator are actually divergent in coordinate space if $B_+ \neq 0$. For instance, the correction $D^{K,\text{ad}}_{n}(\vec{x}|0,0)$ to the Keldysh propagator with $t_1=t_2=0$ in the adiabatic case which comes from $n_p^{\text{ad}}$ (\ref{DensAd}) has the following large $\Lambda$ asymptotic:
\begin{equation}
    D^{K,\text{ad}}_{n}(\vec{x}|0,0) \approx \dfrac{(2\hat{a}\Lambda)^{1/4}}{2\pi^{3/2} |\vec{x}|^{7/4}} |B_++B_-|^2|B_+|^2|B_-|^2 \sin\left[2\sqrt{2\Lambda \hat{a}|\vec{x}|} - \frac{3\pi}{4} \right].
\end{equation}
This expression is evidently divergent in $\Lambda$. Besides that, unlike the usual propagator, it is not exponentially suppressed when $|\vec{x}| \gg \frac{1}{\tilde{m}}$.

As we have noted above, in the CPP the wave frequencies are momentum-independent when $p e^{-\Lambda} \ll 1$, and for large momenta the corrections are suppressed, as we discussed. Hence, similar effect does not occur in CPP. However, it is still interesting to see that in Minkowski space-time even after the resummation of the IR divergences, while they contribute to phases in momentum representation, they are still evident in coordinate representation. Also we see that only the states with $B_+ = 0$ have the well-defined behavior when the mass is changed in the past, and only if the change is adiabatic.

\end{appendices}
\printbibliography
\end{document}